\documentclass{elsart} 
\usepackage{epsfig} 
\usepackage{amssymb} 
\newcommand{\ccdot}{ \hspace{-0.8mm} \cdot \hspace{-0.8mm}} 
\begin{document} 

\begin{frontmatter}

\title{ MORE ABOUT THE COMPARISON  OF~LOCAL~AND~NON-LOCAL 
NN~INTERACTION~MODELS }

\author[Amghar]{A. Amghar}, 
\author[Desplanques]{B. Desplanques} 
\ead{desplanq@isn.in2p3.fr} 
\address[Amghar]{Facult\'e  des Sciences, Universit\'e de Boumerdes, 35000 
Boumerdes, Algeria} 
\address[Desplanques]{Institut des Sciences Nucl\'eaires 
(UMR CNRS/IN2P3--UJF),   F-38026~Grenoble~Cedex,~France }

\begin{abstract}
The effect of non-locality in the NN interaction with an off-energy shell 
character has been studied in the past in relation with the possibility 
that some models could be approximately phase-shifts equivalent. 
This work is extended to a non-locality implying terms that involve an 
anticommutator with the operator $p^2$. It includes both scalar 
and tensor components. The most recent ``high accuracy'' models are 
considered in the analysis. After studying the deuteron wave functions, 
electromagnetic properties of various models are compared with 
the idea that these ones  differ by their non-locality but are 
equivalent up to a unitary transformation. It is found that 
the extra non-local tensor interaction considered in this work tends 
to re-enforce the role of the term considered in previous works, 
allowing one to explain almost completely the difference in the 
deuteron D-state probabilities evidenced by the comparison of 
the Bonn-QB  and Paris models for instance. Conclusions for the 
effect of the non-local scalar interaction are not so clear. 
In many cases, it was found that these terms could explain part 
of the differences that the comparison of predictions for various models 
evidences but cases where they could not were also found. 
Some of these last ones have been analyzed in order to pointing 
out the origin of the failure.
\end{abstract} 

\begin{keyword}
 NN interaction, models, non-locality, deuteron form factors 

\PACS  13.40.Gp, 13.75.Cs, 21.30.Fe 
\end{keyword}
\end{frontmatter}

\section{Introduction}  
A major aim of electron scattering experiment on few-body systems is to 
validate the ingredients entering the description of the process. 
These include the NN interaction models and the associated currents. 
With this respect, the recent measurements of the deuteron form factors 
at JLab \cite{ALEX,ABBO} represent a valuable information. 
The interpretation of these data is not an easy one however. 
There are uncertainties on the nucleon form factors (see Ref. \cite{FORM} 
for a recent review). The effect of relativistic corrections, 
which begin to show up in the explored $Q^2$ range, is approach 
dependent for a part which is not negligeable \cite{RELA}.  
Some genuine effects, due to two-pion exchange, to an underlying 
quark structure of the constituents, to the field-theory foundation 
of the interaction ..., are expected to show up but they are likely 
to be mixed with other ones. Moreover, within a given scheme 
(corresponding to some equation for instance), NN interaction 
models, seemingly different, may represent the same physics. 
This involves different non-localities as expected from phase-shift 
equivalent potentials obtained by unitary transformations.  

As is well known, within a non-relativistic approach, it is always 
possible to describe the NN interaction below the meson-production 
threshold by a local potential for each partial wave separately. 
However, theoretically, the interaction is expected to be non-local. 
This observation motivated a recent study on the role of the non-locality, 
essentially on a phenomenological basis \cite{DOLE}.  The non-locality in 
interaction models has many sources and often results from accounting 
for a theoretical ingredient, explaining that it is not always emphasized 
by the authors. Part of it originates from relativistic corrections 
($\sqrt{m/e}$ factors, structure of Dirac spinors). It has been considered in 
Bonn models \cite{BONN} and is sometimes approximated by linearly 
$p^2/m^2$ dependent terms which makes them tractable in configuration space 
(Paris \cite{PARI}, Bonn-R \cite{BONN} and Nijmegen \cite{STOK}  models). 
Another part comes from the coupling of the  dominant 
NN channel to other ones, $\Delta \Delta$ or $\pi NN$ components. In these 
cases, the original time non-locality is transformed into a spatial one. This 
non-locality is not explicitly present in the models (except of course for the 
energy-dependent ones, full-Bonn for instance) but may be accounted for in the 
fitted parameters. Finally, the compositeness of the nucleon or the exchanged 
meson can also produce some non-locality. It is explicit in quark based 
calculations  (see Ref. \cite{PARI2} for a recent  work together with earlier 
references) but, most often, as in the previous case, it is hidden in the 
parametrization of the potential model.

While different non-localities are present in various models, it is not clear 
whether it helps in describing the NN interaction. These local and non-local 
models may be as many different models, in which case discriminating between 
them is important. One cannot exclude however that they represent  the same 
physics for some part. In this case, there is not so much to study at the NN 
level and, most probably, one should look at other observables. In the example 
of form factors, a preferred model would require less corrections from two-body 
currents beside the genuine ones. The possibility 
to have different models representing the same physics was discussed in the past 
by Friar \cite{FRIA,FRIA1}. For a long time, it was just a curiosity till it was 
realized that more accurate models, namely Paris \cite{PARI} and Bonn-QB 
\cite{BONN}, could realize to some extent this conjecture \cite{DESP1,AMGH1}. 
Terms under consideration have an off-shell character and mainly concern the 
tensor component of the interaction. When these models were used with the 
corresponding currents, it turned out that the discrepancies between the 
predictions for electromagnetic observables tend to decrease. Some was 
left however at the same time that some difference in the deuteron D-state 
probability was unexplained. It was mentioned in Refs. \cite{DESP1,AMGH1} 
that some role could be played by terms which have the form 
of an anticommutator of the kinetic energy with a tensor operator, 
$p^2\, \tilde{W}_T +\tilde{W}_T \, p^2 $. Interestingly, Von Geramb et al. 
\cite{GERA}, 
using inverse scattering methods, were able to derive local equivalent 
potentials for the two Paris and Bonn-QB models, evidencing very 
close  deuteron D-state probabilities, suggesting that the above 
term could also be important in making the predictions of the 
two models closer to each other. These results largely motivate 
the present paper. 

Besides non-local contributions of the type, $i\,(p^2\, U -U\,p^2) $, 
which have been studied in a previous paper and later on by de Forest 
\cite{FORE}, we therefore consider non-local contributions of the 
type, $p^2\, \tilde{W} +\tilde{W}\,p^2 $. 
As in the other case, $W$ can have a scalar as well as a tensor component. The 
first one may originate from overall relativistic $\sqrt{m/e}$ 
normalization factors \cite{ELST}, mostly ignored in non-relativistic 
approaches. Due to the cancellation between the attractive and the repulsive 
components of the force due to $\sigma \;(2\pi)$ and $\omega$ exchanges, its 
effect is not expected to be very large. A larger effect may be due to the 
different Lorentz structures of these couplings, implying that the $p^2/m^2$  
corrections to each of them add together instead of cancelling. The effect, 
which in relativistic mean-field approaches tends to decrease the $\sigma$ 
exchange contribution has a net repulsion. It is part of many models 
\cite{PARI,BONN,STOK} and we believe that it is theoretically relevant. Its 
size is not well known however and, as far as we can see, there was no 
systematic attempt until now to look for an optimized strength.  Perhaps, a 
reason for this is that 
such a term can always be transformed away by a unitary transformation as we 
recalled. If it has some relevance, one should expect that  a smaller number 
of parameters be necessary to fit the same body of data. An indication in this 
sense is given by the comparison of the two Nijmegen models, Nij1 and Nij2 
\cite{STOK}. 
Approximately based on the same philosophy (vertex form factors, ...), but 
differing by a linearly $p^2$ dependent term, they reproduce equally well NN 
scattering data ($\chi^2 \;{\rm per \;datum}= 1.03$). The first one, which is 
non-local, requires however less parameters (41 instead of 47). The non-local 
component with a tensor character is primarily due to the pion exchange. 
This is a well established one as far as the long-range force is concerned.   

When studying the effect of non-local terms of the form, 
$p^2\, \tilde{W} +\tilde{W}\,p^2 $, we will consider a transformation 
up to first order in the interaction. Contrary to 
off-shell effects generated by the terms, $i\,(p^2\, U -U\,p^2) $, the 
determination of the  unitary transformation at the lowest order is not 
straightforward. It requires to solve an integral equation, which can 
only be performed numerically. When this is done, it is possible 
to calculate the effect on observables and see whether such terms 
can explain discrepancies between models evidencing different 
non-localities. This concerns the comparison between models. For our 
purpose, it is not necessary to look at the corrections 
to the interaction model which result from eliminating some non-locality. 
We assume that the fit of the model to scattering data has accounted 
for them, at least to a good approximation. We nevertheless notice 
that such corrections were explicitly considered in the past (see Ref. 
\cite{FRIA3} for a recent work). Ultimately, one is 
interested in a comparison with experiment. With this 
last respect, the previous comparison may be useful to discriminate between 
effects that are simply due to the mathematical representation of the model and 
those due to genuine aspects  such as a two-pion-exchange contribution 
\cite{BONN,PARI}, the choice of the strong $\pi NN $ coupling  
\cite{DESWART,ERIC}, ...

The plan of the paper will be as follows. In the second section, we present the 
non-local terms considered in this work and show how they can be transformed 
away by a unitary transformation at the lowest order. This involves two terms 
with a scalar and a tensor character. The effect on wave 
functions is discussed in the third section for the deuteron case, including   
both its S and D state components. The fourth section is devoted to the effect 
on the deuteron form factors and the structure functions, $A(Q^2)$ and $B(Q^2)$ 
as well as the tensor polarization, $T_{20}(Q^2)$. This part incorporates  
effects 
with an off-shell character described in an earlier work. Comparison of 
predictions for different interaction models is made when correcting them for a 
difference in their non-locality. A few cases where the discrepancy between 
predictions is not explained, and sometimes enhanced, are examined. A discussion 
and a conclusion are given in the fifth section. 
\section{Formalism}

Non-locality can be produced by various contributions to the interaction. That 
one we are considering here originates from an expansion of the spinors entering 
the meson-exchange interaction. The part, which is by far the best identified, 
is due to the pion exchange. Its center of mass expression, up to first order 
terms in ${\bf p}^2/ M^2$, is given by:

\begin{eqnarray} 
V^{\pi}({\bf p'},{\bf p})  = 
-g_{\pi NN}^2 K_{\pi NN}^2 ({\bf k}^2) \; 
\frac{\vec{\tau_1} \ccdot \vec{\tau_2}}{\mu_{\pi}^2 + {\bf k}^2} 
\;\hspace{5.5cm}
\nonumber \\  \times \left( 
\frac{\vec{\sigma_1} \ccdot {\bf k} \; \vec{\sigma_2} \ccdot {\bf k}}{4M^2}-
\frac{\vec{\sigma_1} \ccdot {\bf k} \; \vec{\sigma_2} \ccdot {\bf k}}{16M^4} 
\; ({\bf p'}^2+{\bf p}^2)
\right. \hspace{5.5cm} \nonumber \\ \left. 
 -\frac{  \vec{\sigma_1} \ccdot {\bf (p'-p)} \; 
 \vec{\sigma_2} \ccdot {\bf (p'+p)} +
          \vec{\sigma_2} \ccdot {\bf (p'-p)} \;
           \vec{\sigma_1} \ccdot {\bf (p'+p)} 
  }{32M^4} \; 
({\bf p'}^2-{\bf p}^2)  \right). \label{2e1}
\end{eqnarray}
The above equation  may be written as: 
\begin{equation}
V^\pi=V_S^\pi+V_T^\pi+\{\frac{\bf p^2}{M}, \tilde{W}_S^\pi\}
+\{\frac{\bf p^2}{M}, \tilde{W}_T^\pi \}
+[\frac{\bf p^2}{M}, i\,U^\pi], \label{2e2}
\end{equation}
It can be extended to the exchange of other mesons:
\begin{equation}
V=V_S+V_T+ \{\frac{\bf p^2}{M}, \tilde{W}_S\}
+ \{\frac{\bf p^2}{M}, \tilde{W}_T\}
+[\frac{\bf p^2}{M}, i\,U]. \label{2e3}
\end{equation}
In the above expression, $\tilde{W}_S$ and $\tilde{W}_T$ can be expanded 
in such a way to exhibit their spin-isospin structure. In configuration space, 
they would read:
\begin{eqnarray}
\tilde{W}_S &=& \sum_{s,t} W_S^{s,i}(r) \;P_{s,i} \; ,
\nonumber \\
\tilde{W}_T &=& \sum_t W_T^i(r) \;S_{12}(\hat{r}) \;P_{1,i} \; ,
\end{eqnarray}
where the operators $P_{s,i}$ represent the projectors on different spin and 
isospin channels. The spin tensor operator, shortly written $S_{12}$, is defined 
as $S_{12}(\hat{r})= \vec{\sigma_1} \cdot \hat{r} \; \vec{\sigma_2}
\cdot \hat{r}- \vec{\sigma_1} \cdot \vec{\sigma_2}/3$. Together with 
this definition, we write the deuteron wave function as 
$|\psi(r)>\;\propto [u(r)+3\,w(r)/\sqrt{8}\;S_{12}(\hat{r})]\;|^3S_1>$. 
Working with 
states of given spin and isospin, these projectors can be omitted 
as well as the corresponding indices. The spin-tensor operator cannot 
be however. Without loss of generality, we can therefore make the 
replacements:
\begin{equation}
\tilde{W}_S \leftrightarrow W_S \;(\,=W_S(r)\;),
\;\;\;\tilde{W}_T \leftrightarrow W_T\,S_{12}\;(\,= W_T(r)\;S_{12}(\hat{r})\;).
\end{equation}

The two first terms in Eq. (\ref{2e3}), $V_S$ and $V_T$, 
represent the standard scalar and tensor parts of the non-relativistic   
interaction. These terms are local and are present in all interaction
models. They differ by the precise value of the coupling constant, 
$g_{\pi NN}$, and the expression of the hadronic form factor. This one is 
taken as 1 like in the Reid models or assumes a more realistic form 
like in most other models. The potentials that only retain the two 
first terms at the r.h.s. of Eqs. (\ref{2e2}, \ref{2e3}) are considered as 
local ones. They in particular include the so-called RSC\cite{REID},  
Reid93, Nij2 \cite{STOK} and Argonne V18 \cite{WIRI} models that will be evoked 
in this work. 
They are denoted here as type-I models. 

The so-called Paris \cite{PARI}, Nij1 and Nij93 \cite{STOK} models 
contain a minimal non-locality given 
by a term like the third one in Eq. (\ref{2e3}). Partly justified by its 
appearance at the first order in ${\bf p}^2/ M^2$, the limitation to this term 
is also motivated by the fact that it can be dealt with relatively easily in 
configuration space. These models, like the type-I ones, are generally built 
in configuration space. They are denoted as type-II models. Interestingly, these 
models tend to show a suppression of the deuteron S-wave at short distances 
smaller than local models corresponding to a similar description of the short 
range part of the interaction. An open question is to know whether this can be 
related to the non-local term proportional to $\tilde{W}_S$ in Eq. (\ref{2e3}). 

A last class of models (denoted type-III) is represented by the Bonn ones, QA, 
QB, QC, CD \cite{BONN}. Contrary to the models discussed above, they consider 
the full 
structure of the Dirac spinors describing on mass-shell nucleons. Due to this, 
these models contain all kinds of non-locality present in Eq. (\ref{2e2}) for 
the simplest pion-exchange case and in Eq. (\ref{2e3}) for the most general 
case. They in particular involve terms that have an off-shell character and were 
studied in earlier works \cite{DESP1,AMGH1,FORE}. The dominant contribution was 
provided by the tensor part of $U^{\pi}$ which could explain 2/3 of the 
difference in the deuteron D-state propabilities evidenced by the comparison of 
calculations performed with the Paris and Bonn-QB models. An open question here, 
indirectly answered by von Geramb \cite{GERA}, is to know whether the fourth 
term in Eq. (\ref{2e3}),  $\propto \tilde{W}_T$, can explain the remaining 
part. 

In comparing models, it is important that they are physically equivalent 
with respect to NN scattering properties. The so-called ``high accuracy'' 
models, Nij1, Nij2, Argonne V18 and Bonn-CD should therefore retain our 
attention. Concerning effects in relation with the tensor part of the 
interaction, we will also include the Paris and Bonn-QB models which 
were used in a previous work. These two models provide values for 
the mixing angle, $\epsilon_1$, which are relatively close to 
each other on the one hand, and close to the experimental ones on the other. 

While the good quality of the fit to NN scattering data is essential 
to avoid biases in the analysis we are going to perform, the 
failure to establish some relationship between the various in principle 
phase-shift equivalent models can have different origin. Beyond higher 
order corrections, it may reveal a genuine physical effect. Although 
it can be largely accounted for in the model fitting, the two-pion 
exchange contribution explicitly incorporated in the Paris model 
could be an example of such an effect. It may also reveal that some 
ingredient  in a model is, on the contrary, physically irrelevant. 
Moreover, but not independently, it may suggest that models are 
not completely determined by scattering data in the range where 
they are available. Looking for possible failures can thus 
provide information on the underlying physics, which, after all, is  the 
ultimate goal for studying the NN interaction properties. Anticipating on the 
following, this aspect will mainly involve its scalar part. In the next 
subsections, we show how non-local terms present in Eq. (\ref{2e3}) can be 
transformed away at the lowest order in ${\bf p}^2/ M^2$. This is done 
successively for the scalar and tensor components of the interaction. We will 
concentrate on the terms that have an anticommutator character,  $\{{\bf 
p^2}/M, \tilde{W}_S\}$ and $\{{\bf p^2}/M,  \tilde{W}_T\}$ in Eq. 
(\ref{2e3}). The other 
terms of the commutator type,  $[{\bf p^2}/M,i\, U]$, which are relatively 
easy to deal with, were discussed elsewhere \cite{DESP1,AMGH1,FORE}. 

For definiteness, we denote by $H$ and $H'$ the two Hamiltonians that, in the 
ideal case, would be related to each other by a unitary transformation:
\begin{equation}
H' =e^S\; H \;e^{S^{+}},
\label{2ee1}
\end{equation}
where $S=-S^{+}$.
The Hamiltonian, $H$, will be that one that contains a non-local term, 
$V_{NL}$ which we want to transform away by making a unitary transformation and  
$H'$ will be the resulting Hamiltonian. The two Hamiltonians therefore assume 
the following expressions:
\begin{eqnarray}
H&=&\frac{p^2}{M}+V+V_{NL}. 
\label{2ee3}
 \\
H'&=&\frac{p^2}{M}+V',
\label{2ee2}
\end{eqnarray}
When the non-local term, $V_{NL}$, is transformed away to make it local, the 
interaction $V$ acquires an extra contribution $\Delta\,V$, defined as:
\begin{equation}
\Delta\,V=V'-V.
\label{2ee4}
\end{equation}
Starting from a two-body interaction, $\Delta\,V$ will generally involve 
two-body but also many-body contributions. As to the solutions of the 
Hamiltonians, $H$ and $H'$, they are denoted $\Psi$ and $\Psi'$.
 
\subsection{Non-locality in the scalar part of the interaction models}
In this subsection, we consider two models that differ by a non-local term 
proportional to $W_S$. The non-local term, $V_{NL}$, therefore assumes the 
following expression:
\begin{equation}
V_{NL}=\{\frac{\bf p^2}{M}, W_S(r)\}. \label{2e5}
\end{equation}
The physical origin of the last term in Eq. (\ref{2e5}) has been briefly 
mentioned in the introduction. The first candidate for such a contribution is 
provided by the examination of the $\omega$ and $\sigma$ (2$\,\pi$  in a 
S-state) exchanges. The cancellation that occurs for these contributions at the 
zeroth order in $p/M$ is absent at the order $(p/M)^2$, hence a possible 
sizeable  term. In comparison, the contribution due to the pion exchange in Eq. 
(\ref{2e2}) contains extra factors of the order $(p/M)^2$. Due to the above 
cancellation, minimal relativity factors $\sqrt{m/e}$, whose effect was 
studied in a first step by Elster et al. \cite{ELST}, may  only be a part of  
the dominant contribution. Non-locality could also arise from eliminating 
degrees of freedom involving mesons in flight or baryons excitations. It however 
appears that the corresponding spatial and time non-localities, individually 
large, tend to cancel each other. This probably explains the success of a local 
potential in reproducing the spectrum of the Bethe-Salpeter equation 
\cite{AMGH2}. 

Equation (\ref{2ee1}) together with the expression of $V_{NL}$ given by Eq. 
(\ref{2e5}) cannot be solved easily but taking into account that the term we 
want to remove is of order $(p/M)^2$, we can solve it order by order by making 
an expansion of the unitary transformation that we denote $e^{S_S}$. At the 
lowest order in the interaction, we have thus to fulfill the following equation:
\begin{equation}
\{\frac{\bf p^2}{M}, W_S\}+[S_S,\frac{{\bf p^2}}{M}]= \Delta\,V_S^0 \; , 
\label{2e7}
\end{equation}
where $\Delta\,V_S^0$ must be local. 
This is achieved by choosing $S_S$ as:
\begin{equation}
S_S=i\;\Big(\frac{\vec{p} \ccdot \vec{r} \; V_0(r)+V_0(r) \; \vec{r} \ccdot 
\vec{p}}{2}\Big)\; , \label{2e8}
\end{equation}
where the function $V_0$ satisfies the following first-order differential 
equation:
\begin{equation}
 V_0(r)+\vec{r} \ccdot \vec{\nabla}V_0(r)=W_S(r) \; . \label{2e9}
\end{equation}
The solution of this equation is given by the following integral:
\begin{equation}
V_0(r)=-\frac{1}{r}\int_r^\infty W_S(r')dr'. \label{2e10}
\end{equation}
The integral can be calculated numerically provided that the integrand is not 
too singular. The singular character of $V_0$ due to the presence of the front 
factor $1/r$ is partly removed by the presence of the factor, $\vec{r}$, in the 
expression of $S_S$,  Eq. (\ref{2e8}). It is likely that this singular character 
is the consequence of limiting ourselves  to the lowest order terms in both 
expanding $e^{S_S}$ and in the interaction. It should be removed in a complete 
calculation but this supposes a fully non-local unitary 
transformation. 

While the non-local term in Eq. (\ref{2e3}) is removed at the lowest order, an 
extra local contribution to the interaction with same order in the coupling 
constant is generated. Denoted $\Delta V_S^0$ in Eq. (\ref{2e7}),  its 
expression is given in the appendix. Let's precise here that its locality holds 
for each partial wave separately, as expected from phase-shift equivalent 
models. On the other hand, it is somewhat more singular than the interaction 
models we started from or contains terms with extra powers of $1/r$. By fitting 
the model parameters to the scattering NN data, such terms may be partly 
accounted for but we do not expect to completely explain the discrepancy between 
different models, the new terms having a functional form that is generally 
discarded in building models. Finally, an important difference is to be noticed  
with the treatment of the non-local term of the commutator type,  $[{\bf 
p^2}/M,i\, U]$, 
made elsewhere \cite{DESP1,AMGH1}. The correction, $\Delta{V}$, contains 
in the present case a contribution of the first order in the coupling $g^2$, 
$\Delta{V}^0$, while it was of the second order for this other term.

For practical applications, we will consider the two potential models, Nij1 
and Nij2, which have been built with approximately the same ingredients, 
except evidently for the presence of a non-local term in the first one, 
which has the form of the term considered in this work. 
We also considered other couples of models, Paris and RSC for instance. 
Their probing strength as to their possible unitary equivalence was not 
found very strong. Instead, the comparison pointed out to a discrepancy 
having its source in the different values for the quantities, $A_S$ 
and $a_t$ (and possibly the ratio $\eta=A_D/A_S$) obtained with these 
models. Some correlation between $A_S$ and $a_t$ is well known 
but a correlation to $A_D/A_S$, which appears to be supported by the 
examination of various models, is not so. Characterizing the 
asymptotic behavior, the difference evidenced by the comparison 
of these values that are observable ones, cannot be removed 
by the unitary transformation considered here.

\subsection{Non-locality in the tensor part of the interaction models.}
In this subsection, we consider two models that differ by a non-local tensor 
term of the anticommutator type:
\begin{equation}
V_{NL}=\{\frac{\bf p^2}{M},\,W_T(r)\, S_{12}(\hat{r}) \}.
\label{2e1212}
\end{equation}
The non-local tensor term of the commutator type has been discussed at 
length in earlier works \cite{DESP1,AMGH1}. 

To eliminate the non-local term under consideration, we proceed as in the 
previous subsection, with the same approximations (expansion of the unitary 
transformation, now denoted $e^{S_T}$, up to first order in $S_T$ and 
first-order 
terms in the interaction retained). The quantity $S_T$  has to be chosen in such 
a way that its commutator with the kinetic energy term  ${\bf p^2}/M$ cancels 
the term $\{ {\bf p^2}/M,\,W_T\,S_{12} \}$ without generating the same terms 
as those we want to remove. This implies to solve an equation similar to Eq. 
(\ref{2e7}) for the scalar case:
\begin{equation}
\{\frac{\bf p^2}{M},\,W_T \,S_{12}\}+[S_T,\,\frac{{\bf p^2}}{M}]
= \Delta\,V_T^0 \; , 
\label{2e7bis}
\end{equation}
where $\Delta\,V_T^0$ has to be local. 
This supposes to take two terms in  $S_T$ and is achieved by: 
\begin{eqnarray}
S_T=\frac{i}{2} \Bigg(\vec{p} \ccdot \vec{r} \;\; S_{12}(\hat{r}) \; V_1(r) 
\hspace{7cm}  \nonumber \\     
+\Big(\vec{\sigma_1} \ccdot \vec{p} \; \vec{\sigma_2} \ccdot \vec{r} +
     \vec{\sigma_2} \ccdot \vec{p} \; \vec{\sigma_1} \ccdot \vec{r} - 
\frac{2}{3} \;
     \vec{\sigma_1} \ccdot \vec{\sigma_2} \;\vec{p} \ccdot \vec{r} \Big) \; 
V_2(r)
+  h.c. \Bigg),
\label{2e11}
\end{eqnarray}
The functions, $V_1(r) $ and $V_2(r)$, have 
to satisfy the following relations:
\begin{eqnarray}
r\,(V'_1(r)+V'_2(r))&=&W_T(r),\label{2e12} \\ 
V_1(r)+V_2(r)+(rV_2(r))'&=&0.
\label{2e13}
\end{eqnarray}
Despite it is more complicated than Eq. (\ref{2e9}), the above system of 
equations can be formally solved with the result:
\begin{eqnarray}
 V_2(r)&=&-\frac{1}{r}\int_r^\infty dr'\int_{r'}^\infty 
\frac{W_T(r'')}{r''}dr'',
\label{2e14}  \\
V_1(r)&=&-V_2(r)-\frac{1}{r}\int_r^\infty \frac{W_T(r')}{r'}dr'.
\label{2e15}
\end{eqnarray}
Numerical values for $V_1(r) $ and $V_2(r)$ are then obtained by performing 
successively the integrations in Eqs. (\ref{2e14}) and (\ref{2e15}). Concerning 
the singularities entering these expressions, the same observations as for 
$V_0(r) $ can be made here. 

Similarly to the scalar case, the removal of the non-local tensor term of the 
anticommutator type generates an extra contribution to the interaction, 
$\Delta V^0_T$, with a local character. Its expression is given in the appendix. 
Again, it evidences an analytic structure different from what is usually 
included in potential models. A large part of it is certainly accounted for by 
fitting its parameters to scattering data. Whether the remaining 
part can show up is probably difficult to answer in view of other uncertainties.

Applications will be made for the couple of the Paris and Bonn-QB models 
that we considered in a previous work for part of the non-local tensor 
component with a commutator type, assuming implicitly that non-local scalar 
terms of the anticommutator type are more or less the same. We also 
considered the couple of models Argonne V18 and Bonn-CD. In this case, 
a non-local scalar  term is also included, slightly obscuring the analysis.

\section{Comparison of wave functions}

In this section, we compare wave functions obtained from various models 
differing by their non-locality. In principle, and as far as the models are 
phase-shift equivalent, it should be possible to relate the corresponding wave 
functions according to the equality:
\begin{equation}
\Psi'=e^{S} \; \Psi,
\label{3e16}
\end{equation}
where $\Psi$ and $\Psi'$ are solutions of Schr\"odinger equations with 
Hamiltonians respectively given by Eqs. (\ref{2ee3}) and (\ref{2ee2}).

Consistently with our approximation retaining the first term in the expansion of 
the unitary transformation $e^{S}$ made in its determination, we assume that the 
correction  to $\Psi$ produced by the application of this transformation is 
given by 
\begin{equation}
\Delta \Psi= S \; \Psi.
\label{3e17}
\end{equation}
In practice, we will calculate this quantity, add it to $\Psi$  and compare 
the result to the wave function obtained independently in a different model, 
$\Psi'$, which it could be compared with. Expressions of $\Delta \Psi$ in the 
case of a non-local term with an off-shell character, $\propto i\,U$, have been 
given 
in Refs. \cite{DESP1,AMGH1}. We here give these expressions for the other 
non-local terms, $\propto W_S$ and $\propto W_T$. 

\noindent 
{\it Correction to the wave function from transforming away the non-local term  
$\{ {\bf p^2}/M, W_S\}$}\\
For the first case, one gets separately for the $S$ and $D$ wave 
components of the deuteron wave function:
\begin{eqnarray}
 \Delta u=\frac{1}{2}\, W_S\, u +  V_0 \,r \, u',
\label{3e18} \\
\Delta w=\frac{1}{2}\,  W_S\,  w+ V_0 \,r \,  w'. 
\label{3e19}
\end{eqnarray}
Using the expression of $V_0$ in term of $W_S$, Eq. (\ref{2e9}), one can easily 
check that the correction to the norm, always at the first order in $S$, is zero 
separately for the $S$ and $D$ waves:
\begin{equation}
\int_0^\infty dr\;u\,\Delta u=\int_0^\infty dr\;w\,\Delta w=0.
\label{3e20}
\end{equation}
This result is consistent with the unitary character of the transformation and 
the fact it does not mix the two waves.

\noindent 
{\it  Correction to the wave function from transforming away the non-local term  
$\{ {\bf p^2}/M,\,W_T\,S_{12}\}$}\\
Expressions for the case of the term $\propto W_T$ are more complicated:
\begin{eqnarray}
\Delta u&=&\frac{\sqrt{8}}{3} \; \left((V_1+5\,V_2+\frac{1}{2} \, W_T) \, w
+ (V_1+ 2\,V_2)( r \, w'-w)\right),
 \label{3e21} \\
\Delta w&=&\frac{\sqrt{8}}{3}
\left((V_1-V_2+\frac{1}{2}\,W_T)\,u + (V_1+2\,V_2)(r\,u'-u)
 \right) \nonumber \\
& &- \frac{2}{3}  
    \left((V_1+2\,V_2+\frac{1}{2} W_T) \,w  + (V_1+2\,V_2)(r\,w'-w)\right). 
 \label{3e22}
\end{eqnarray}
As previously, using the above expressions, one can check that the correction to 
the norm is zero but the result now holds for the total sum of the $S$ and $D$ 
waves contributions:
\begin{equation}
\int_0^\infty dr\; (u \, \Delta u+w \, \Delta w)=0.
 \label{3e23}
\end{equation}

\begin{figure}[htb]
\begin{center}
\mbox{ \epsfig{ file=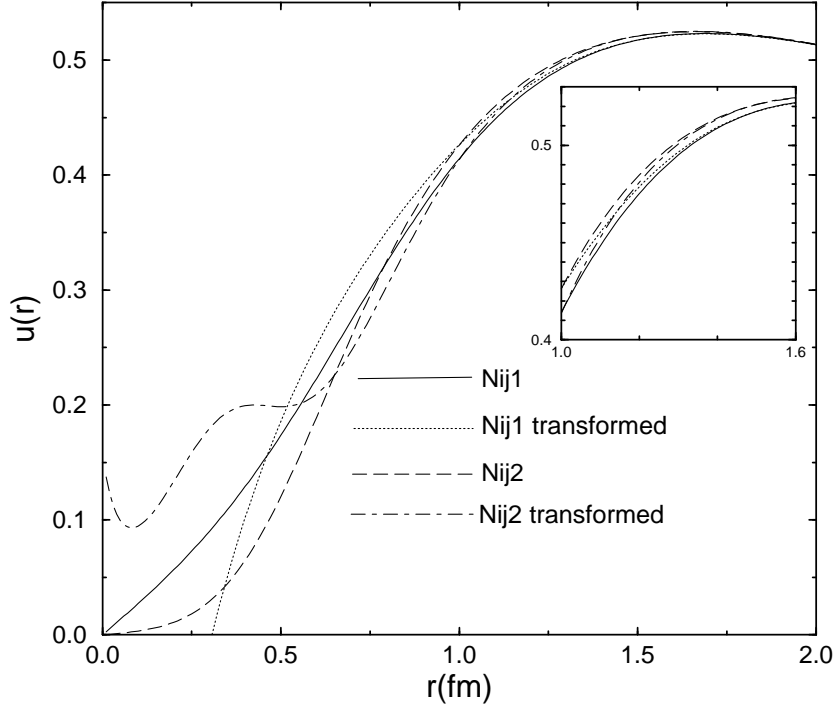, angle=270, width=11cm}}  
\end{center}
\caption{Effect of a non-local scalar term, $\{ {\bf p^2}/M, W_S\}$, on the 
deuteron S-waves:  The effect is shown for the Nij1 and Nij2 models. The 
corresponding deuteron S-waves, $u(r)$, are represented by the continuous and 
dashed lines respectively. They should be compared with the transformed waves 
obtained from the other model, represented in the figure by dashed-dotted and 
dotted lines. The insert shows an enlarged part of the figure in the region 
where the non-locality effect shows up at large distances. For this calculation, 
$W_S$ is taken from the Paris model.}
\label{fig1s}
\end{figure}

\begin{figure}[htb]
\begin{center}
\mbox{ \epsfig{ file=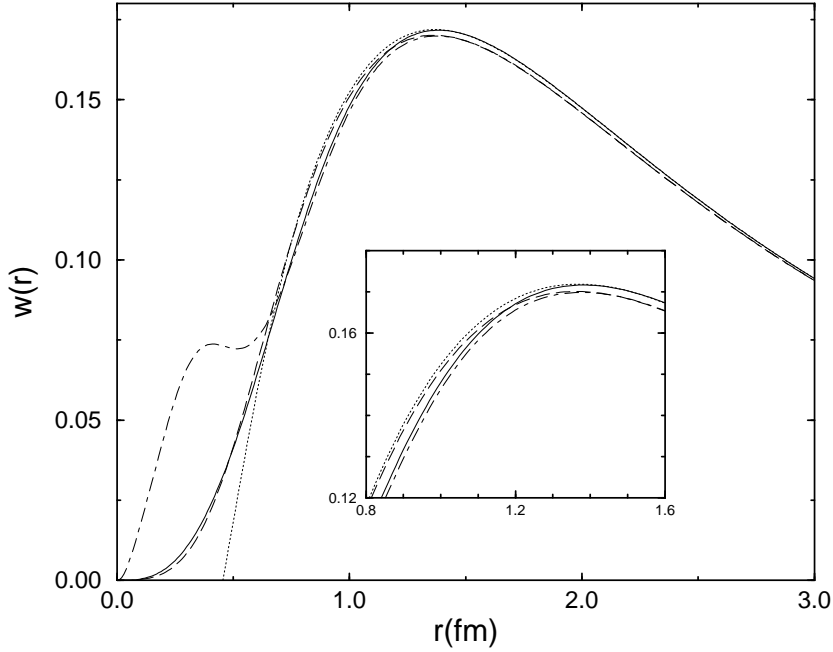, angle=270, width=11cm}}  
\end{center}
\caption{Effect of a non-local scalar term, $\{ {\bf p^2}/M, W_S\}$, on the 
deuteron D-waves: The effect is shown for the Nij1 and Nij2 models. Same as in 
Fig. \ref{fig1s}, but for the deuteron D-waves, $w(r)$.}
\label{fig2d}
\end{figure}

\begin{figure}[htb]
\begin{center}
\mbox{ \epsfig{ file=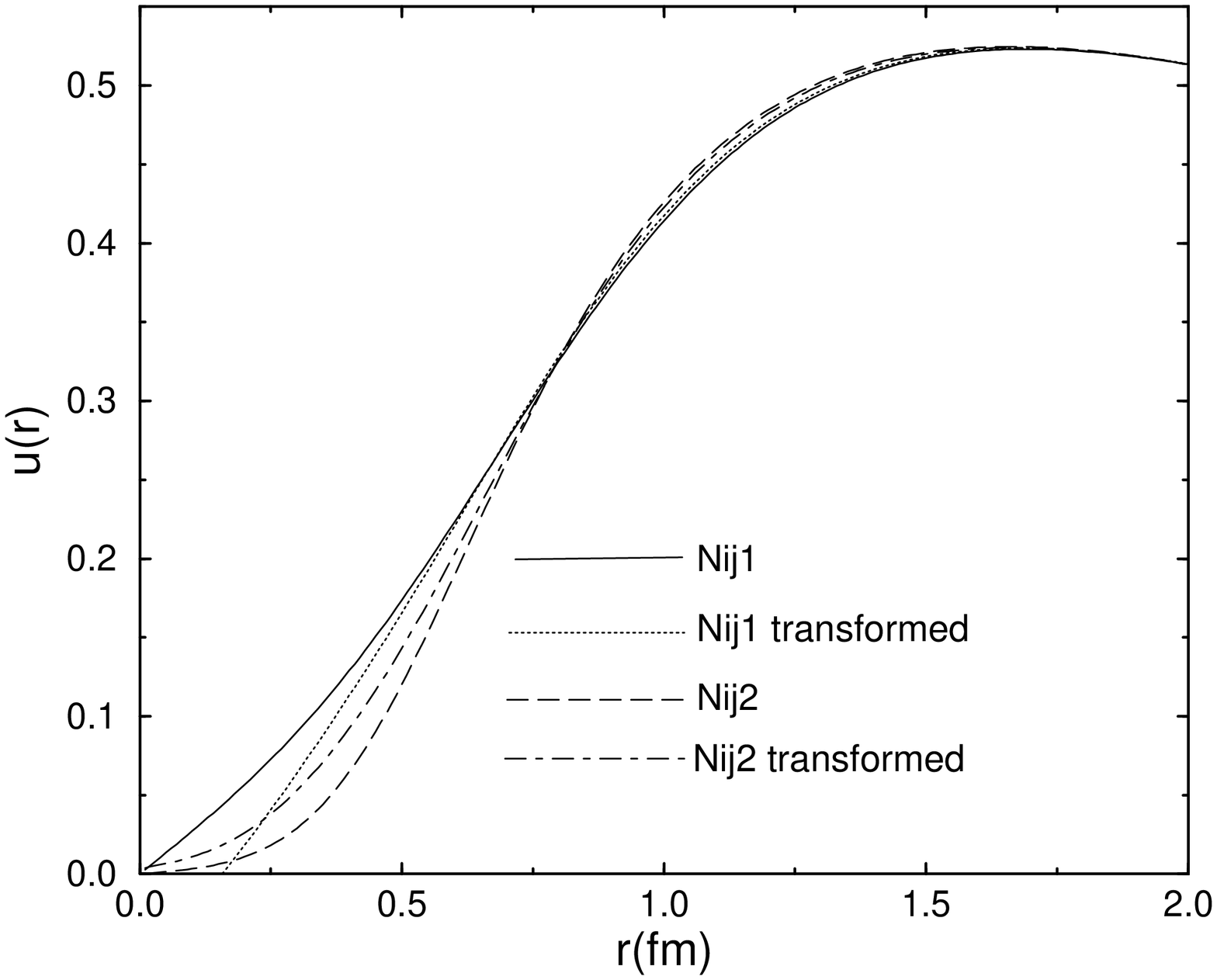, angle=270, width=6cm} \hspace{1cm}\epsfig{ 
file=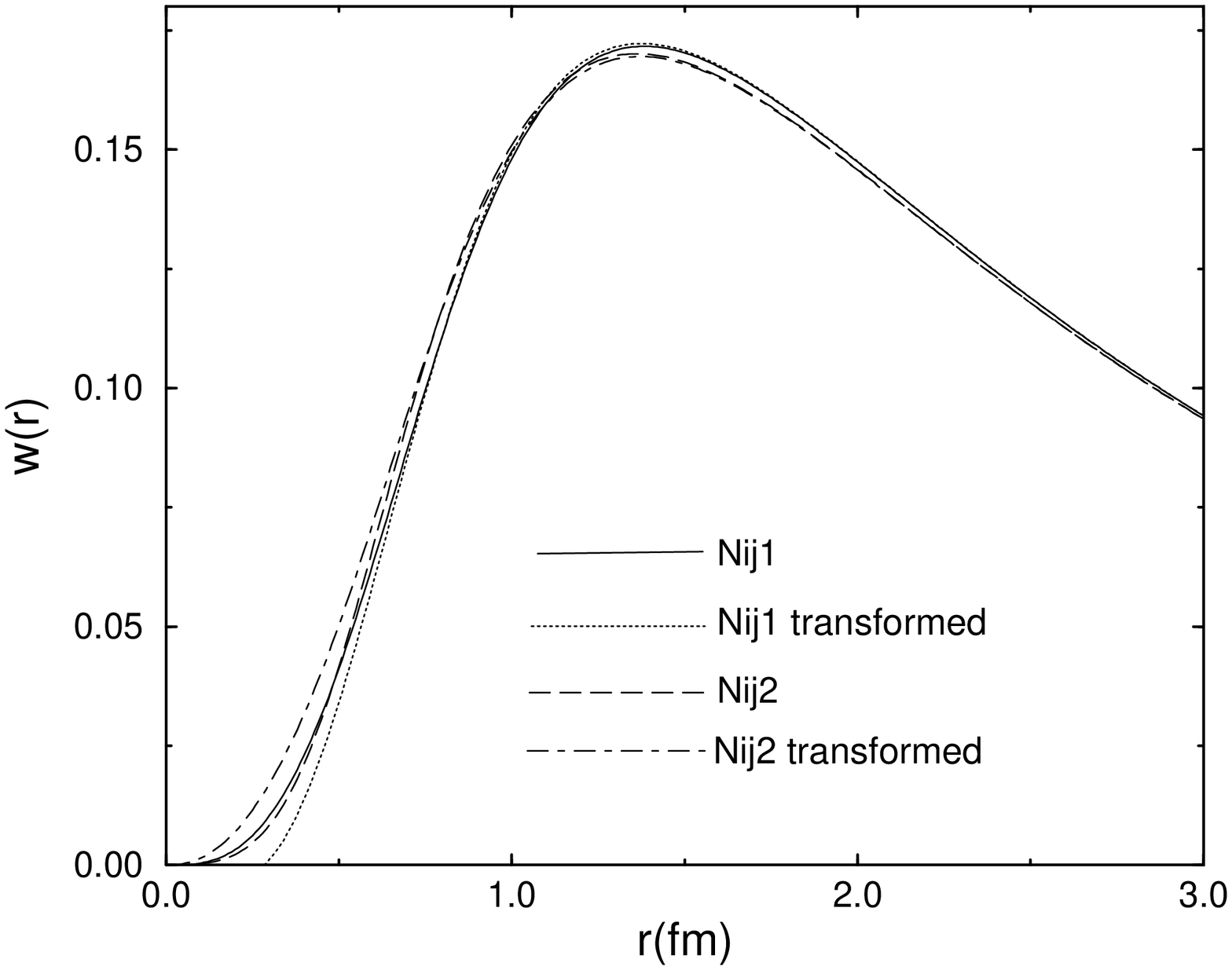, angle=270, width=6cm}}  
\end{center}
\caption{Effect of a non-local scalar term, $\{ {\bf p^2}/M, W_S\}$, on the 
deuteron S- and D-waves: The effect is shown for a different $W_S$ in 
calculating the transformed S- and D- waves from the Nij1 and Nij2 models. 
The corresponding curves are respectively given in the right and left parts; 
same as in Fig. \ref{fig1s} otherwise. For this calculation, $W_S$ is 
taken from a meson-exchange model. }
\label{fig3x}
\end{figure} 

\begin{figure}[htb]
\begin{center}
\mbox{ \epsfig{ file=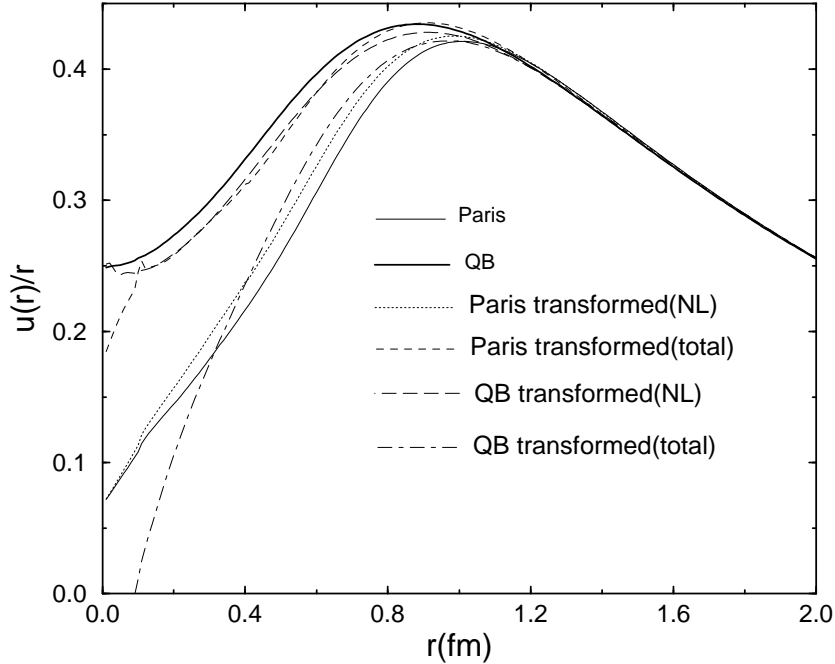, angle=270, width=11cm}}  
\end{center}
\caption{Effect of a non-local tensor term together with off-shell 
contributions: The effect is considered for the S-waves of the Paris and Bonn-QB 
models.  Curves show the quantity $u(r)/r$ to better emphasize the short range 
behavior. The thin and thick continuous lines represent the corresponding 
quantity for these models. The transformed wave functions accounting for the 
non-local tensor term, $\{ {\bf p^2}/M,\,W_T\,S_{12}\}$, are given respectively 
by dotted 
and long-dashed lines (labelled NL). The wave functions accounting for the whole 
effect are given by the short-dashed and dash-dotted lines. These last ones 
should be compared with the wave function of the other model.}
\label{fig3s}
\end{figure}

\begin{figure}[htb]
\begin{center}
\mbox{ \epsfig{ file=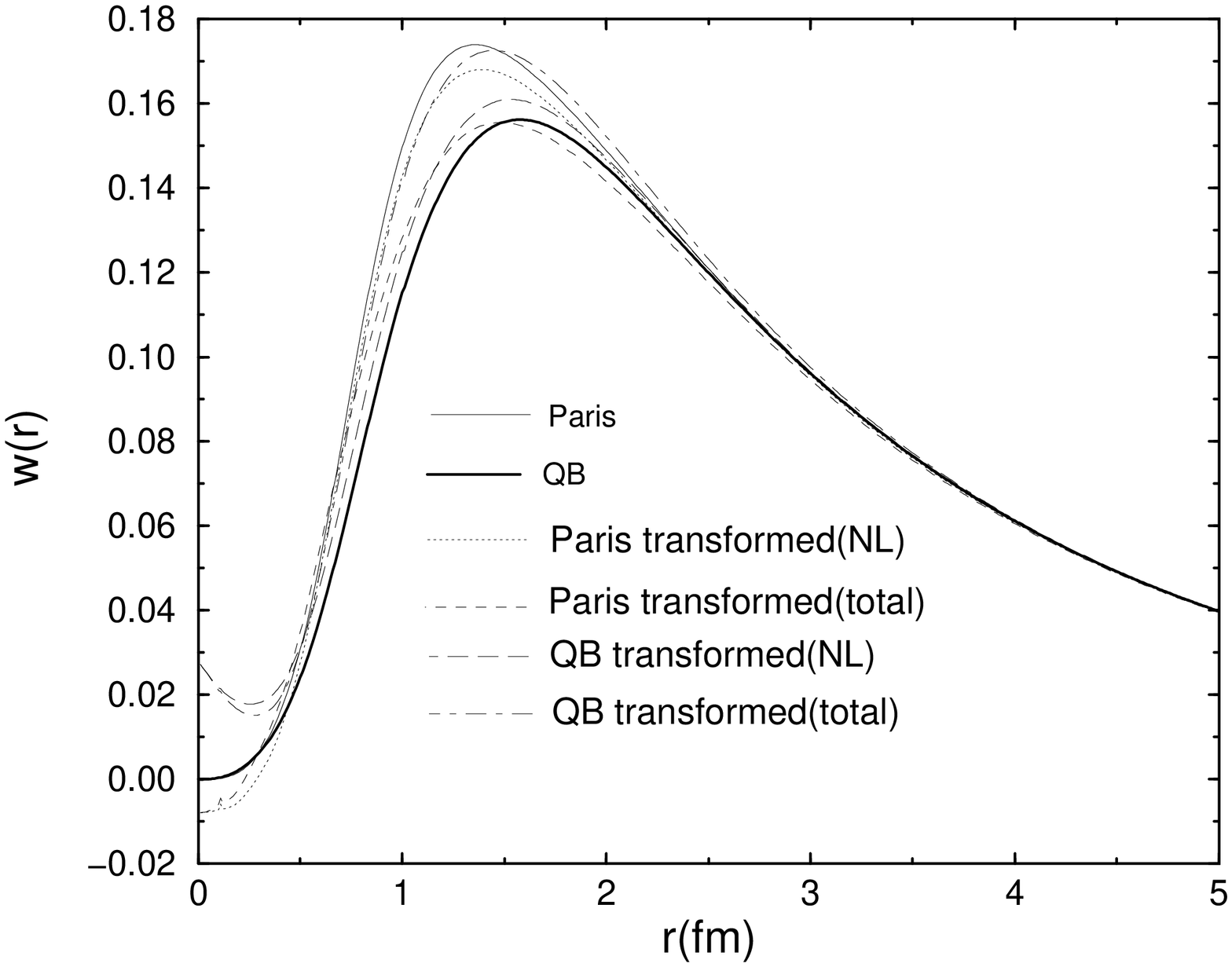, angle=270, width=11cm}}  
\end{center}
\caption{Effect of a non-local tensor term together with off-shell 
contributions: The effect is considered for the D-waves of the Paris and Bonn-QB 
models. Curves represent the usual D-waves, $w(r)$; same  as in Fig. \ref{fig3s} 
otherwise.}
\label{fig4d}
\end{figure}

\noindent 
{\it Comparison of wave functions for Nij1 and Nij2 models}\\
In Figs. \ref{fig1s} and \ref{fig2d}, we show the effect of the 
transformation  discussed above in 
the case of a scalar non-local term, $\{ {\bf p^2}/M, W_S\}$. This is done 
for the couple of models Nij1 and Nij2 that precisely differ by such a term.
As far as the unitary transformation is truncated to the first order term in 
$S$, applying the transformation $e^S$ on $\Psi$ or the inverse transformation 
$e^{-S}$ on $\Psi'$ is not equivalent. We therefore present results where the 
two cases are considered in the same figure. Beyond $0.8\,{\rm fm}$, where 
corrections are small enough to be treated perturbatively, it is found that both 
the S and D waves become closer to the wave functions they should be compared 
with (see the insert). Some 
discrepancy remains at the largest distances. In the D-wave case, it partly 
reflects a difference in the asymptotic normalizations, $A_D$, that are not 
affected by the unitary transformation. In the S-wave case, it is not clear 
whether the remaining difference around $1.4\,{\rm fm}$ is an indirect 
consequence of 
the previous one, of the choice of $W_S$ (actually taken from the Paris model in 
a first step analysis) or simply of the model parametrization. It is noticed 
that the wave functions at these distances are likely to be sensitive to the 
medium range of the interaction, which itself involves scattering data at 
intermediate energies. The effect of a difference in reproducing these data, 
such as the 8\%  one for the $\epsilon_1$ mixing angle around $100\,{\rm MeV}$ 
for the two models considered here, cannot be removed, evidently, by the unitary 
transformation. 

At lower distances, examination of the figures, especially for the S-wave, 
evidences structures. They are due for a large part to the perturbative 
character of our approach. At the lowest distances, the effect of the 
transformation goes in the right direction. It therefore seems reasonable to 
attribute the smoother behavior of the S-wave for some models to their $p^2$ 
dependence. The effect is however too large. It is likely that a more accurate 
treatment would transfer part of the effect at larger distance, around 
$0.6\,{\rm fm}$, where the present effect is too small or even has a sign 
opposite to what was expected.  This transfer is not arbitrary however as the 
norm should remain unchanged.

The existence of the above structures may introduce a bias in the comparison of 
form factors and other observables that will be made in the next section. 
Actually, the effect is limited by the fact that in calculating the form 
factors, we only retain the linear corrections in order to preserve the norm, 
according to Eqs. (\ref{3e20}) and (\ref{3e23}). Moreover, in the integrand, 
this correction is 
multiplied by a wave function which tends to zero in the relevant domain. The 
actual effect on form factors would thus involve a factor $q^2\,r^3/24$ in the 
integrand and is consequently suppressed. It is roughly of the order 
$q^2\,r_0^4/96$ where $r_0$ is of the order of $0.4\,{\rm fm}$. Furthermore, one 
cannot exclude cancellations with corrections at higher $r$. As will be seen in 
the next section, it is what occurs and the different structures should be 
considered consistently.

Due to the theoretical uncertainty on $W_S$, we estimated the transformed wave 
functions with assuming that this quantity was produced by single meson 
exchanges ($\pi, \rho, \omega, \sigma$) with standard values for the parameters. 
Results are presented in Fig. \ref{fig3x}. They evidence smaller structures than 
in Figs. \ref{fig1s} and \ref{fig2d}, probably in relation with a smaller 
strength of the $W_S$ term. By using in the following a term taken from the 
Paris model, it is likely that its strength is not underestimated. 

\noindent 
{\it Comparison of wave functions for Bonn-QB and Paris models}\\
In Figs. \ref{fig3s} and \ref{fig4d}, we show the effect of the 
transformation  discussed above in the case of a tensor non-local term,  
$\{ {\bf p^2}/M,\,W_T\,S_{12}\}$. For this calculation, $ W_T\,S_{12}$ 
is taken as the sum of a $\pi$-exchange contribution appearing in Eq. 
(\ref{2e1}), which provides the dominant contribution, and a 
$\rho$-exchange contribution, which has a cut-off role. Together 
with this non-local term, we consider the effect of the off-shell 
term studied in a previous work. A monopole cut-off  form factor, 
$(\Lambda^2-\mu^2)/(\Lambda^2+k^2)$, with $\Lambda=1.05$  GeV, 
is introduced in all calculations \cite{AMGH1} so that to kill 
the short-range part of the interaction, for which the analysis 
developped in this work is more uncertain, both theoretically and 
mathematically. For the S-wave (Fig. \ref{fig3s}), 
the transformed wave functions are relatively close to the one 
they should be compared with. The effect of the first term, 
$\{ {\bf p^2}/M,\,W_T\,S_{12}\}$, which arises from the action on 
the D-wave (therefore different of the scalar case) goes in the 
right direction. Most of the effect comes from the term with an 
off-shell character. Notice that for this curve, we plotted $u(r)/r$, 
in order to better see the effects at short distances. Quite similar 
results hold for the D-wave shown in Fig. \ref{fig4d}. It is immediate 
from these results that when considering the electromagnetic properties 
of the deuteron, the predictions of the different models will tend 
to be close to each other when the currents that are associated to 
each model are accounted for. Concerning the little structures 
appearing in the transformed wave functions, the remarks made previously 
for the comparison of Nij1 and Nij2 wave functions could be repeated 
here. The amplitude of these structures is much less important here 
however.

\section{Effect on deuteron form factors and electromagnetic observables}
Given a physical system and an underlying theory, it can be described by an 
infinity of interaction models, depending on the choice of degrees of freedom. 
This is quite similar to the choice of a frame or a basis in making an 
experiment or a calculation. This freedom is at the origin of the non-observable 
character of the wave function. Concerning the system itself, this is realized 
by using phase-shift equivalent potentials. As for the interaction with an 
external probe, described by an operator denoted $O$,  the equivalence of the 
models implies a 
relation such as:
\begin{equation}
\langle \Psi | O | \Psi \rangle=\langle \Psi' | O' |\Psi' \rangle.
\label{4e24}
\end{equation}
In this expression, the operator $O'$, which can contain in the minimal case 
one-body terms but also many-body ones, will here refer to the electromagnetic 
interaction. Using the expression relating  $\Psi'$ and $\Psi$, Eq. 
(\ref{3e16}), it is found that the above equality supposes that:
\begin{eqnarray}
 O' &= & e^{S} \; O \; e^{-S}= O + \Delta \, O ,
\nonumber \\ 
\Delta \, O &=  & [S,O]+\dots.
\label{4e25}
\end{eqnarray}
In the case where $O$ is a one-body operator, it is immediately found that  
the correction at the r.h.s. has in the minimal case a two-body character. 
In order to make a relevant comparison of predictions of observables from 
different representations, one has therefore to account for the contribution due 
to $\Delta \, O $ when calculations are performed. This is what we are doing in 
this section. The question of making a comparison to measurements, which in any 
case supposes to consider the contribution of other currents, is evoked in the 
conclusion. 

For the off-shell correction considered in Refs. \cite{DESP1,AMGH1}, the 
commutator $\Delta \, O $ could be calculated explicitly, leading to two-body 
currents that, except for a factor, had for some part a structure identical to 
the pair-term contribution to the charge operator. The expression for the 
magnetic operator could not be related to any known two-body operator. A similar 
situation holds for the transformation considered here for terms of the form 
 $\{ {\bf p^2}/M, \tilde{W}_{S,T}\}$. One thus gets the following contribution 
for the 
isoscalar charge operator in configuration space at the lowest order :
\begin{eqnarray}
[S, O] &=&\frac{i}{2}\; e^{i\,\vec{q} \cdot \vec{r}/2} \;\Bigg( 
\vec{q} \ccdot \vec{r} \;\;  V_0(r) + \vec{q} \ccdot \vec{r} \; S_{12}(\hat{r}) 
\; V_1(r) 
\nonumber\\ & & \hspace{1.8cm}
+  \Big(  \vec{\sigma_1} \ccdot \vec{q} \; \vec{\sigma_2} \ccdot \vec{r} +
          \vec{\sigma_2} \ccdot \vec{q} \; \vec{\sigma_1} \ccdot \vec{r} - 
\frac{2}{3} \;
     \vec{\sigma_1} \ccdot \vec{\sigma_2} \;\vec{q} \ccdot \vec{r} \Big) \; 
V_2(r) \Bigg).  
\label{4e26}
\end{eqnarray}

In practice, calculations can be performed indistinctively in two ways. 
The first one, in the spirit of equivalent models, consists in calculating 
the matrix element of $O+\Delta \, O $ with the wave function, $\Psi'$, 
and compare it to the matrix element calculated with $O$ together with 
the wave function,  $\Psi$. In the second one,  the first matrix element 
is replaced by the matrix element of $O$ with the transformed  wave 
function, $e^{-S}\;\Psi'$. While in this later approach conclusions 
immediately stem from the examination of transformed w.f. discussed in the 
previous section, the former approach is not without interest. Indeed, some 
of the corrections to charge, quadrupole and magnetic form factors (for the 
scalar part $W_S$) have a relatively simple expression:
\begin{eqnarray}
\Delta F_C(Q^2)=G_E^S(Q^2)\int_0^\infty dr\;xj_1(x)\;   
\Bigg( \Big(u^2(r)+w^2(r)\Big)\;V_0(r) \hspace{1.5cm} \nonumber \\ 
 +\frac{2\sqrt{8}}{3}
\Big(u(r)\;w(r)-\frac{1}{\sqrt{8}} \; w^2(r)\Big)\;
\Big(V_1(r)+2\,V_2(r)\Big)\Bigg) , 
\nonumber\\
\Delta F_Q(Q^2)= \frac{6\sqrt{2}}{Q^2} \; G_E^S(Q^2)
\int_0^\infty  dr \hspace{6cm} \nonumber \\
\times \Bigg( \Big(3\,j_2(x)-xj_1(x)\Big)\; 
\Big(u(r)\;w(r)-\frac{1}{\sqrt{8}}\;w^2(r)\Big)  \; V_0(r) \hspace{1cm}
 \nonumber \\ 
+\sqrt{8}\;   j_2(x) \;
\Big(-u^2(r) +w^2(r)+\frac{1}{\sqrt{2}} \,u(r) \; w(r)\Big) \;V_2(r)  
\hspace{1.3cm}
 \nonumber \\ 
+\frac{\sqrt{2}}{3} \;\Big(3\,j_2(x)-xj_1(x)\Big)  \; \hspace{5cm}
 \nonumber \hspace{1cm}\\ 
\times \Big(u^2(r)+\frac{3}{2}\; w^2(r)-\sqrt{2} \; u(r) \; w(r)\Big) \; 
\Big(V_1(r)+2\,V_2(r)\Big)    \Bigg),  
\nonumber\\
\Delta F_M(Q^2)
 =\frac{3}{4} \; G_E^S(Q^2)\int_0^\infty dr\;3\,j_2(x)\;  w^2(r)\;V_0(r) 
\hspace{3.5cm}\nonumber\\ 
+G_M^S(Q^2)\int_0^\infty dr\;\Bigg(xj_1(x)\;\Big(u^2(r)-\frac{1}{2}w^2(r)\Big)
\hspace{2.5cm}  \nonumber\\  
+\frac{1}{2}\;\Big(3\,j_2(x)-xj_1(x)\Big)\;\Big(w^2(r)+
\sqrt{2}\;u(r)\;w(r)\Big)\Bigg)
\;V_0 (r)+...
\label{4e27}
\end{eqnarray}
where $x=Q\,r/2$.  The above expressions can be also obtained from those using 
the transformed wave functions by performing an integral by parts and taking 
into account that the term $....|_0^\infty$ is zero. Contrary to these ones 
however, the integrand is quite smooth and does not evidence structures at short 
distances. Far to suggest numerical inaccuracies or errors in the formalism, the 
structures that we discussed in the previous section have to be taken as they 
are and, most important, dealt with consistently. The dots at the last line of 
the expression for the magnetic form factor contain terms with the derivatives 
of the wave functions. Unlike the charge and quadrupole form factors, these ones 
cannot be removed using an integration by parts.

When comparing observables, we will proceed in two ways, depending on which 
model predictions are compared with. This is dictated by the first order 
approximation made in expanding the unitary operator $e^S$, which is required 
for consistency, as explained above, but prevents to keep the unitary  character 
of the transformation. In assessing the two sets of results, it is likely that 
the truth should be in between. We successively consider the static observables, 
the form factors and finish the section by mentioning a few peculiar results.

\subsection{Static observables}
\noindent
{\it Effect of  $\{ {\bf p^2}/M, W_S\}$}\\
When comparing predictions for models differing by a term $\{ {\bf p^2}/M, 
W_S\}$, we found very small effects for the static observables, $r^2_D$ and 
$Q_D$, and obviously none for $\mu_D$ (in relation in this case with the 
absence of change of $P_D$). The non-local term being short-range-behaved, 
we rather expect effects in the dynamical observables (form factors) 
as it can be inferred from the discussion of the wave functions 
in Sect. 3. Concerning the models, Nij1 and Nij2, we looked at more 
specifically, static observables are quite close to each other, 
leaving little room for possible effects. The difference in the values 
of $r^2_D$, $0.004\,{\rm fm}^2$, can be traced back to a difference in 
the asymptotic normalizations $A_S$ for 75\% (models roughly fulfill the 
relation $\Delta r_m\simeq 1.9 \,\Delta A_S$). What we calculated,  
$\simeq 0.001\,{\rm fm}^2$, is consistent with the unexplained part. 
For the quadrupole moment, we found a correction that has a sign 
opposite to what would be required by the comparison of the two models. 
Again, the major part of the difference in this case is likely 
to have another origin (see our discussion of the D wave around 
$1.4\,{\rm fm}$ in Fig. \ref{fig2d}, Sect. 3). As for the the 
magnetic moment, it is unchanged since the normalizations 
of the S- and D-waves are separately unchanged.

\noindent
{\it Effect of  $\{ {\bf p^2}/M,\,W_T\,S_{12}\}$ and other terms}\\
Effects of  the  tensor term, $\{ {\bf p^2}/M,\,W_T\,S_{12}\}$, are 
significantly more important than for the scalar one. In a previous work 
\cite{AMGH1}, we could explain 2/3 
of the difference in the D-state probabilities for the Bonn-QB and Paris 
models (0.78\%) by considering a non-local tensor term with an off-shell 
character, $[{\bf p^2}/M,\,i\,U]$. By adding the above term, we can now explain 
a 
difference of 0.74\%
(0.14\% for the new term alone). 
The deuteron D-wave probability enters in the magnetic moment according to the 
relation $\Delta \mu_D= 3/2\;(\mu_S-1/2) \;\Delta P_D$. The difference for the 
two 
models is $\Delta \mu_D$ is $0.0045\;{\rm nm}$ while we calculate $0.0043\;{\rm 
nm}$. Another observable sensitive to the tensor force is the quadrupole moment. 
Considering the quantity $Q_D/(A_S\;A_D)$, that may be less sensitive to 
differences in $A_S$ or $A_D$ left unchanged by the unitary transformation we 
are dealing with, we found a difference of $0.19\,{\rm fm}^3$ or $0.20\,{\rm 
fm}^3$ depending on which model is corrected. The difference in the predictions 
is 
$0.23\,{\rm fm}^3$. Finally, we explain 60\% of the difference in the  squared 
charge radius evidenced by the comparison of the two models, Bonn-QB and Paris. 
This last observable is directly related to the slope of the deuteron structure 
function, $A(Q^2)$. 

As far as static observables are concerned, it thus appears that they are very 
close to each other once the effects of their supposed unitary equivalence are 
taken into 
account.

\subsection{Comparison of form factors with models Nij1 and Nij2, effect of the 
non-local scalar term,  $\{ {\bf p^2}/M, W_S\}$ }

The effect of the non-local scalar term,  $\{ {\bf p^2}/M, W_S\}$, on 
electromagnetic observables is shown in three figures for the case of the Nij1 
and Nij2 potentials models, which differ by such a term. Here and below, we 
expect that if this term has some relevance in explaining the difference in the 
wave functions of these models, then predictions for the observables should be 
closer to each other when the appropriate currents are used. The ratio of these 
predictions, that we represent in the figures, should therefore approach 1.
\begin{figure}[htb]
\begin{center}
\mbox{ \epsfig{ file=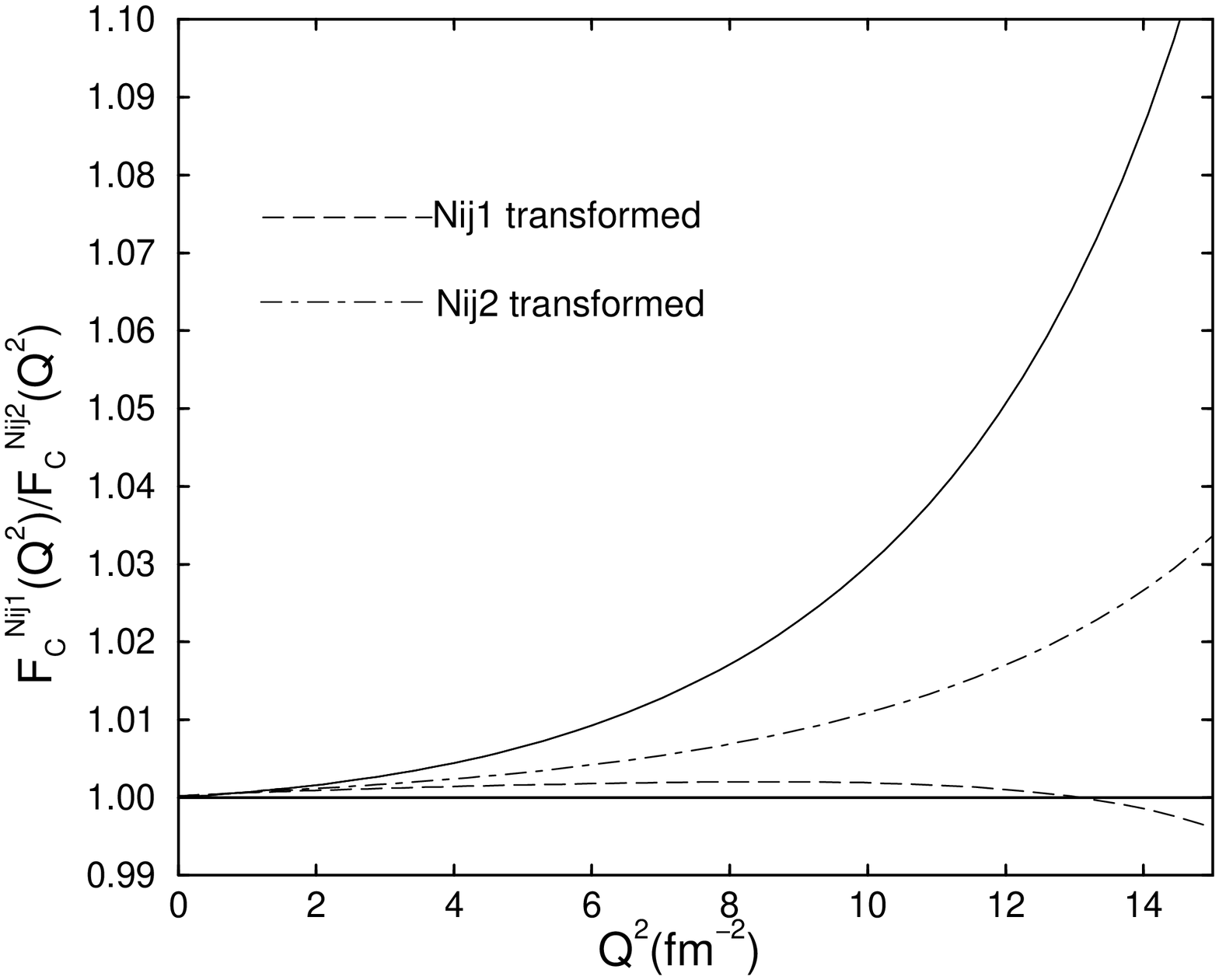, angle=270, width=6cm} \hspace{1cm}\epsfig{ 
file=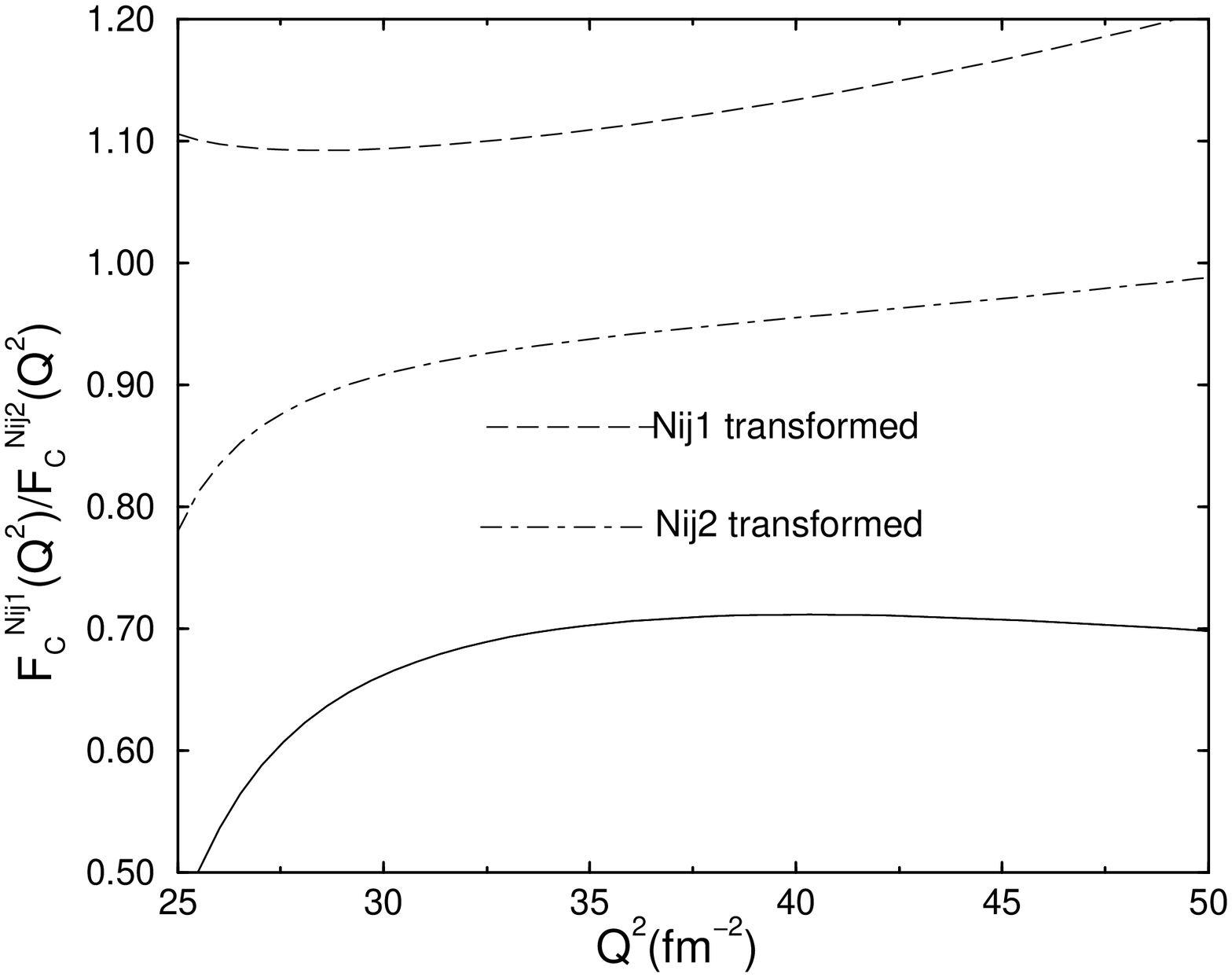, angle=270, width=6cm}}  
\end{center}
\caption{Effect of a non-local term, $\{ {\bf p^2}/M, W_S\}$, on  charge form 
factors: Curves represent the ratio of the Nij1 to the Nij2 charge form factors 
from  $Q^2=0$ to $Q^2=15\,{\rm fm}^{-2}$ for the left part and  from  $Q^2=25$ 
to $Q^2=50\,{\rm fm}^{-2}$ for the right part. The ratio of the uncorrected 
predictions is given by the continuous line. The ratios with the transformed 
Nij1 and Nij2 w.f. (or including the associated two-body currents) are 
respectively given by the dashed and the dash-dotted lines.}
\label{figx}
\end{figure}

\begin{figure}[htb]
\begin{center}
\mbox{ \epsfig{ file=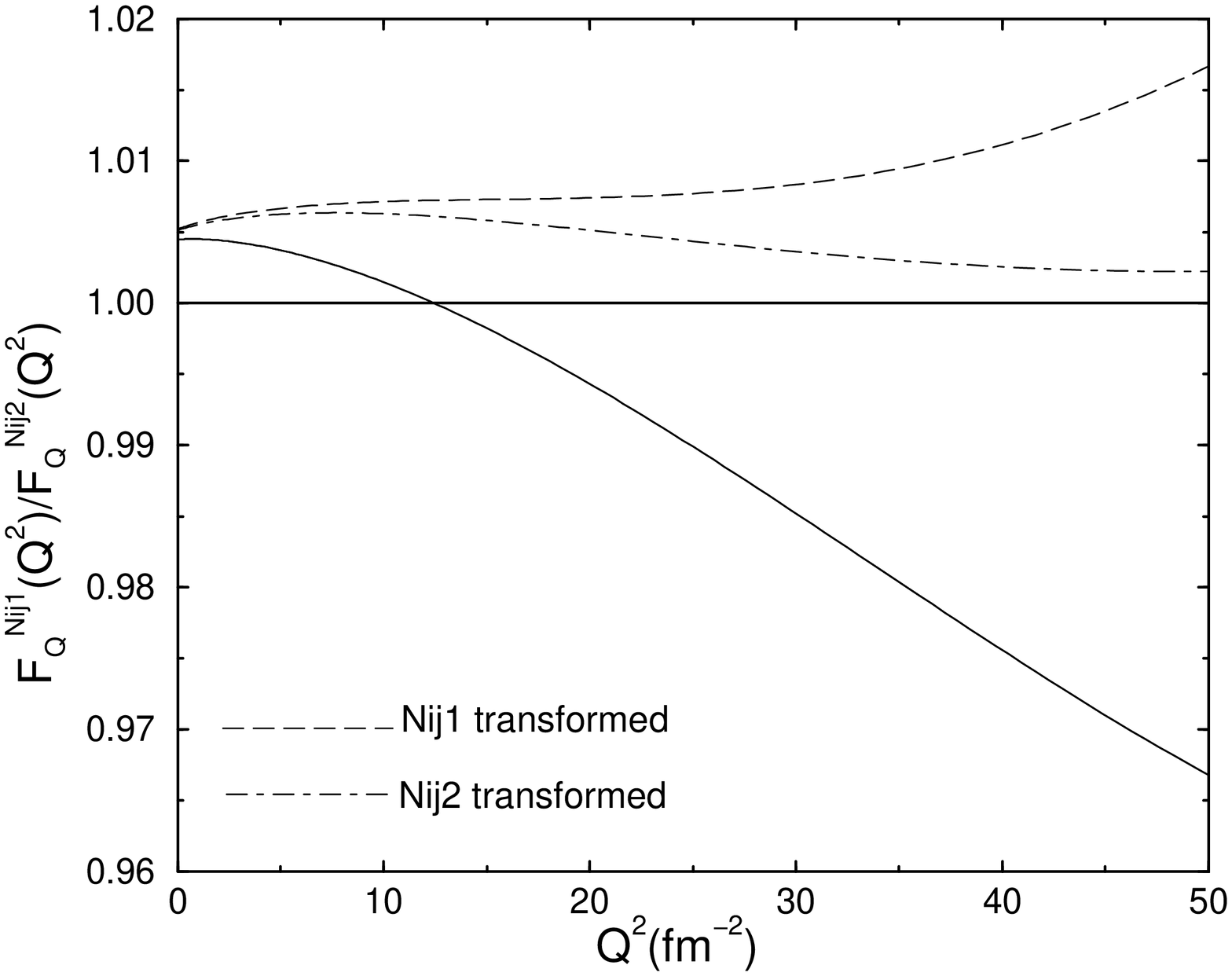, angle=270, width=6cm} \hspace{1cm}\epsfig{ 
file=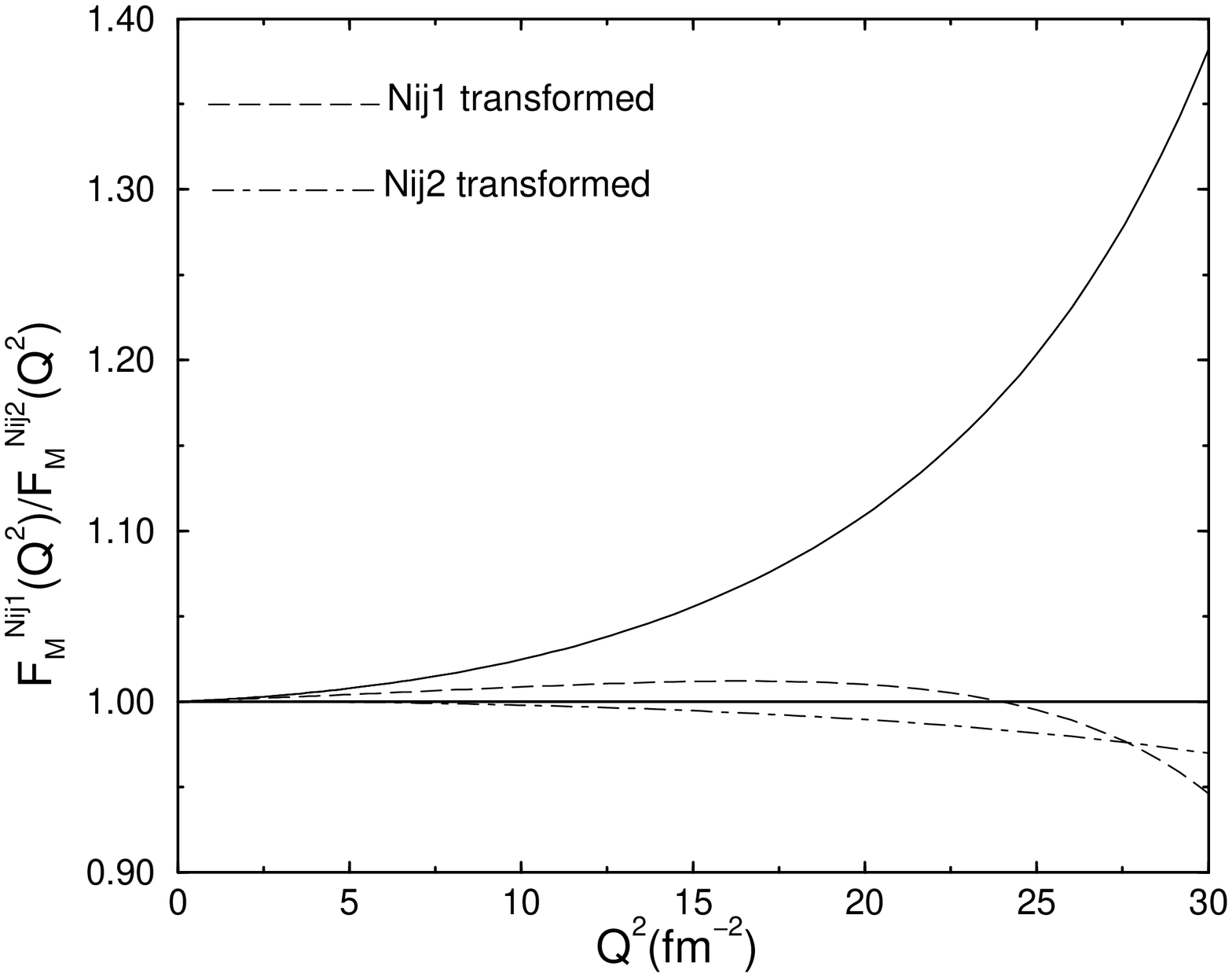, angle=270, width=6cm}}  
\end{center}
\caption{Effect of a non-local term, $\{ {\bf p^2}/M, W_S\}$, on  quadrupole and 
magnetic form factors: Curves represent the ratio of the Nij1 to the Nij2 
quadrupole and magnetic form factor from  $Q^2=0$ to $Q^2=50\,{\rm fm}^{-2}$ for 
the first one (left part) and  $Q^2=0$ to $Q^2=30\,{\rm fm}^{-2}$ for the second 
one (right part). They are defined as in Fig. \ref{figx}.}
\label{figy}
\end{figure} 

\begin{figure}[htb!]
\begin{center}
\mbox{ \epsfig{ file=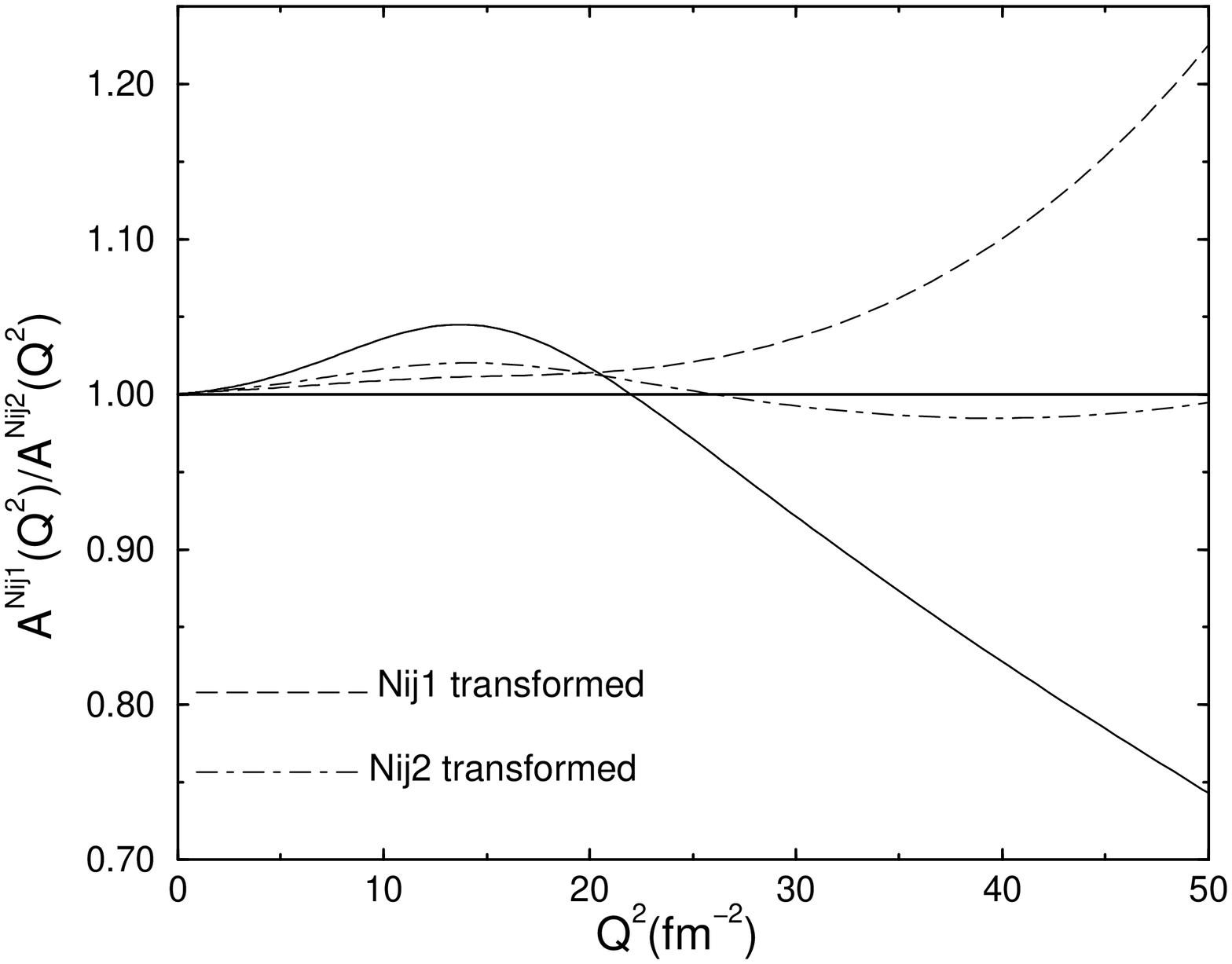, angle=270, width=6cm}  \hspace{1cm} 
\epsfig{ file=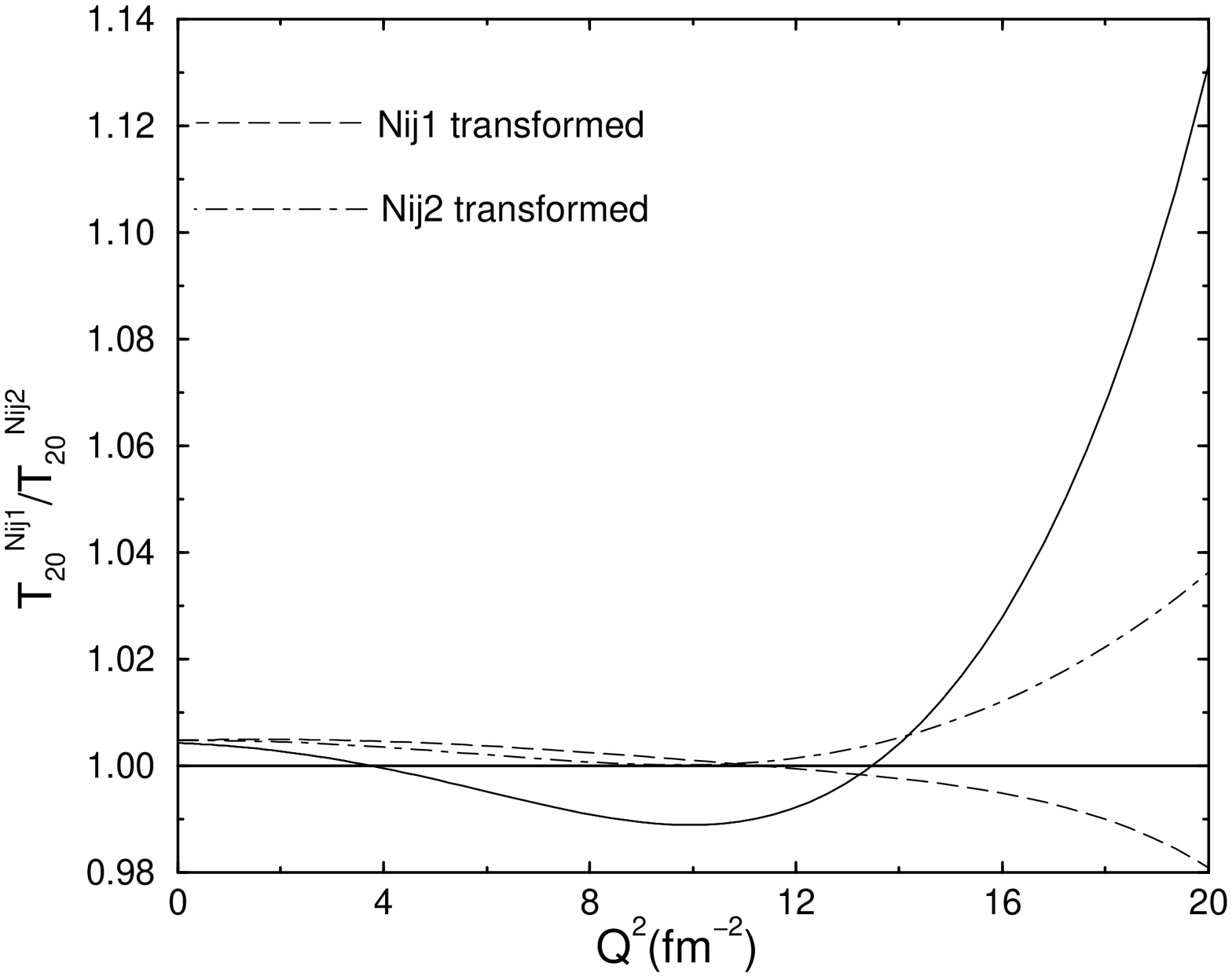, angle=270, width=6cm}    }
\end{center}
\caption{Effect of a non-local term, $\{ {\bf p^2}/M, W_S\}$, on $A(Q^2)$ and 
$T_{20}(Q^2)$ observables: Curves represent the ratio of the  Nij1 to the Nij2 
$A(Q^2)$ and $T_{20}(Q^2)$ observables, respectively for the left and right 
parts. 
They are defined as in Fig. \ref{figx}. }
\label{figAT20}
\end{figure} 

The effect on charge form factors is seen in Fig. \ref{figx} that has been split 
into two parts, due to the appearance of a zero in the range 15-25 fm$^{-2}$. 
While the effect was hardly significant for static observables, examination of 
the figure indicates that it is quite relevant at higher $Q^2$ and goes in the 
right direction. Results do not depend much on whether two-body currents are 
added to one model or to the other.  A similar conclusion holds for the 
quadrupole form factor as well as the magnetic one shown in Fig. \ref{figy}. 
Notice that the last one is given up to 30 fm$^{-2}$ due to the presence of a 
zero at a slightly larger $Q^2$. 

It is also instructive to look at quantities which are closer to what is 
measured, namely the deuteron structure function $A(Q^2)$, and the tensor 
polarization at $\theta_e=70^0$, $T_{20}(Q^2)$. They are given 
in Fig. \ref{figAT20}. Not surprisingly, they nicely confirm the above 
conclusions.

Results quite similar to the above ones have been obtained when 
extended to other interaction models (Nij1 replaced by Paris or Nij93 ; 
Nij2 replaced by Argonne V18 or Reid93). The agreement is especially 
good for those models that better fit the NN scattering data. 
Due to the various uncertainties already mentioned, we will mainly 
retain that the currents associated with the removal of a term,  
$\{ {\bf p^2}/M, W_S\}$, in the interaction can reasonably account for 
differences in model predictions, both in sign and in magnitude. Some 
differences remain however (and hopefully) unexplained. They will be 
discussed at the end of this section.  

\subsection{Comparison of form factors with models Bonn-QB and Paris, 
effect of the non-local tensor term}
\begin{figure}[htb]
\begin{center}
\mbox{ \epsfig{ file=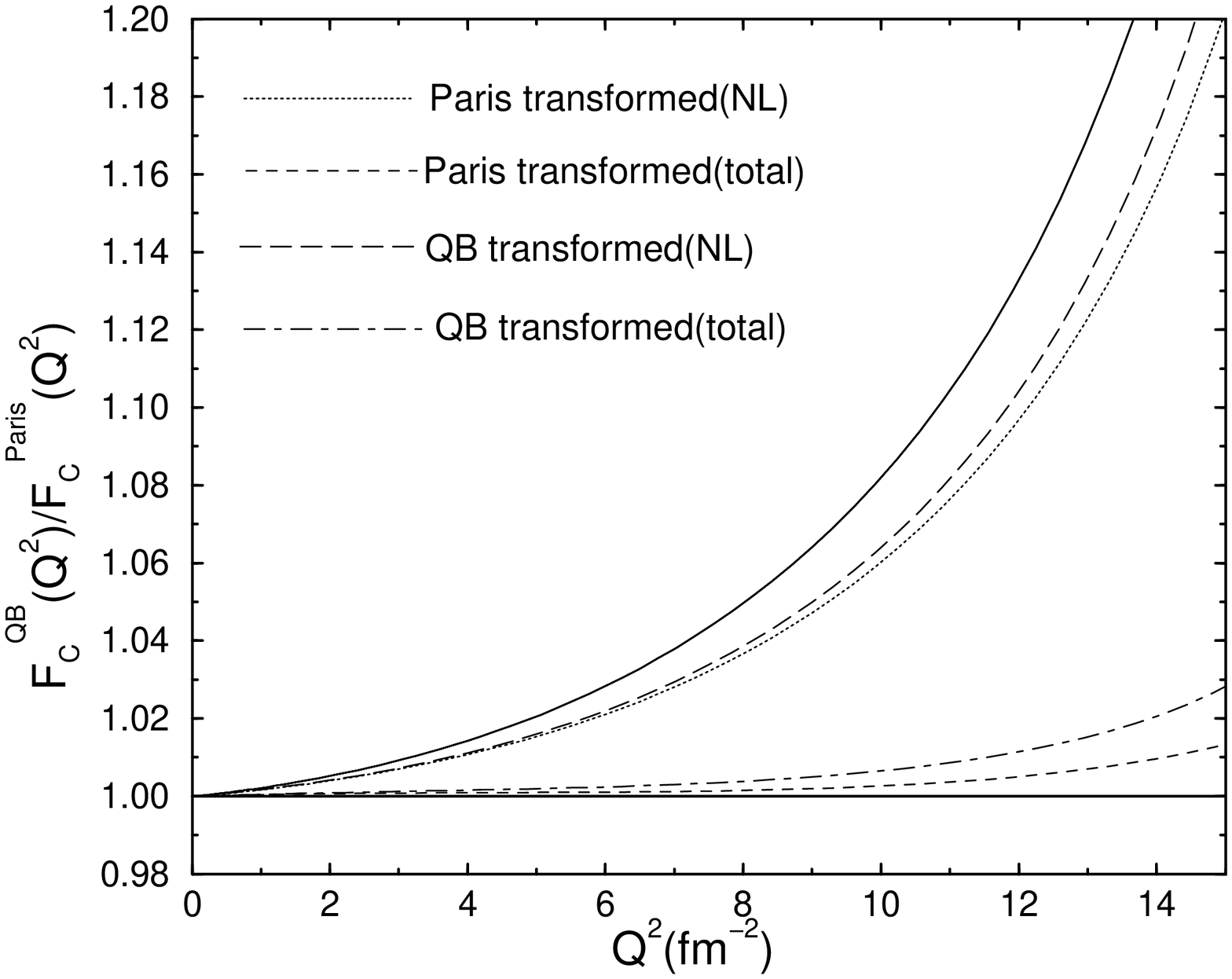, angle=270, width=6cm} \hspace{1cm}
\epsfig{ file=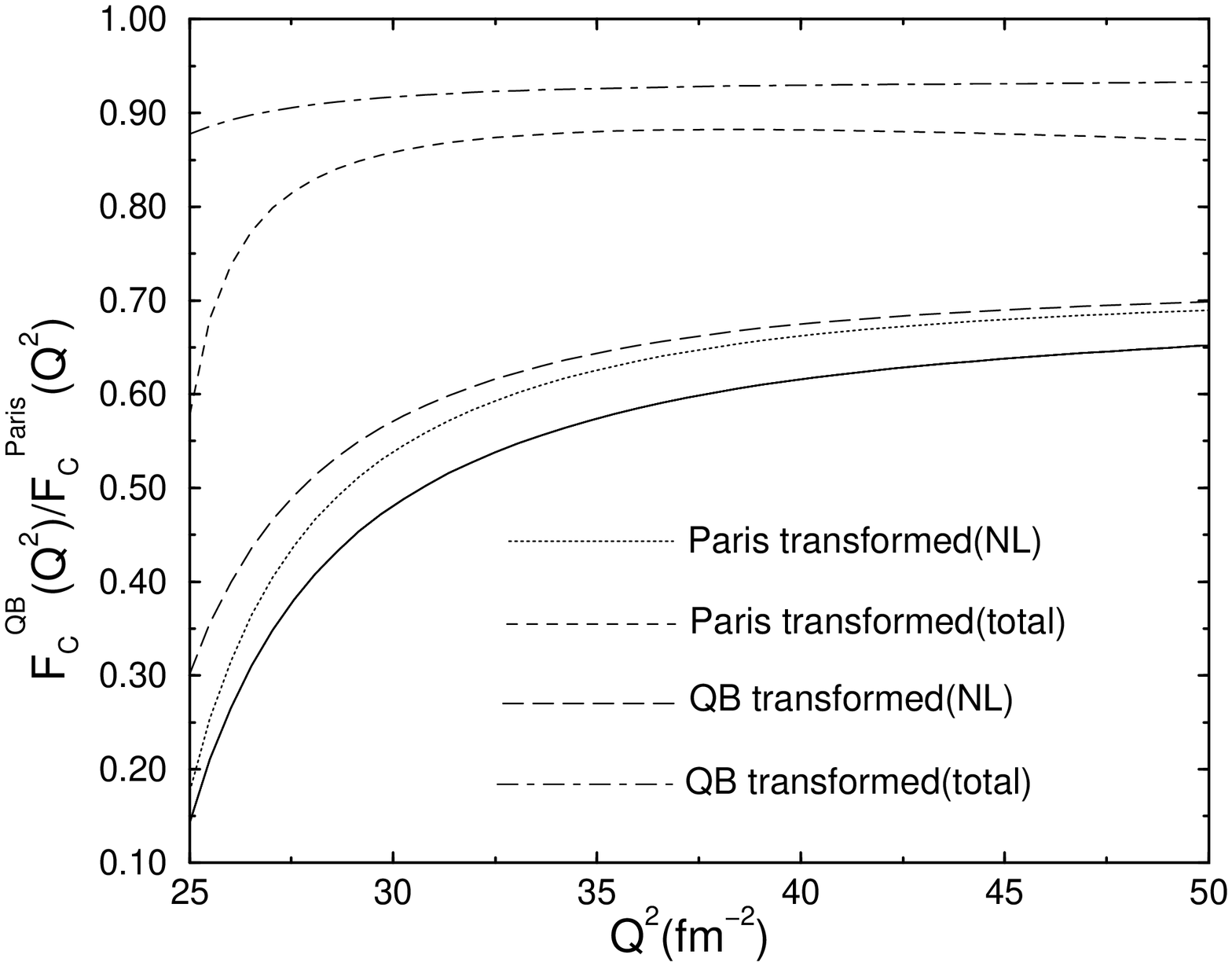, angle=270, width=6cm}}  
\end{center}
\caption{Effect of a non-local term, $\{ {\bf p^2}/M,\,W_T\,S_{12}\}$ together 
with 
off-shell effects for the charge form factor: Curves represent the ratio of 
the Bonn-QB to the Paris charge form factors from  $Q^2=0$ to $Q^2=15\,{\rm 
fm}^{-2}$ for the left part and  from  $Q^2=25$ to $Q^2=50\,{\rm fm}^{-2}$ 
for the right part. The ratio of the uncorrected predictions is given by the 
continuous line. The ratio when the Paris (Bonn-QB) w.f. is transformed for 
the first ( $\{ {\bf p^2}/M,\,W_T\,S_{12} \}$) and second (off-shell) 
non-locality is 
given by the dotted and short-dashed (long-dashed and dash-dotted) lines.}
\label{figx2}
\end{figure}

\begin{figure}[htb]
\begin{center}
\mbox{ \epsfig{ file=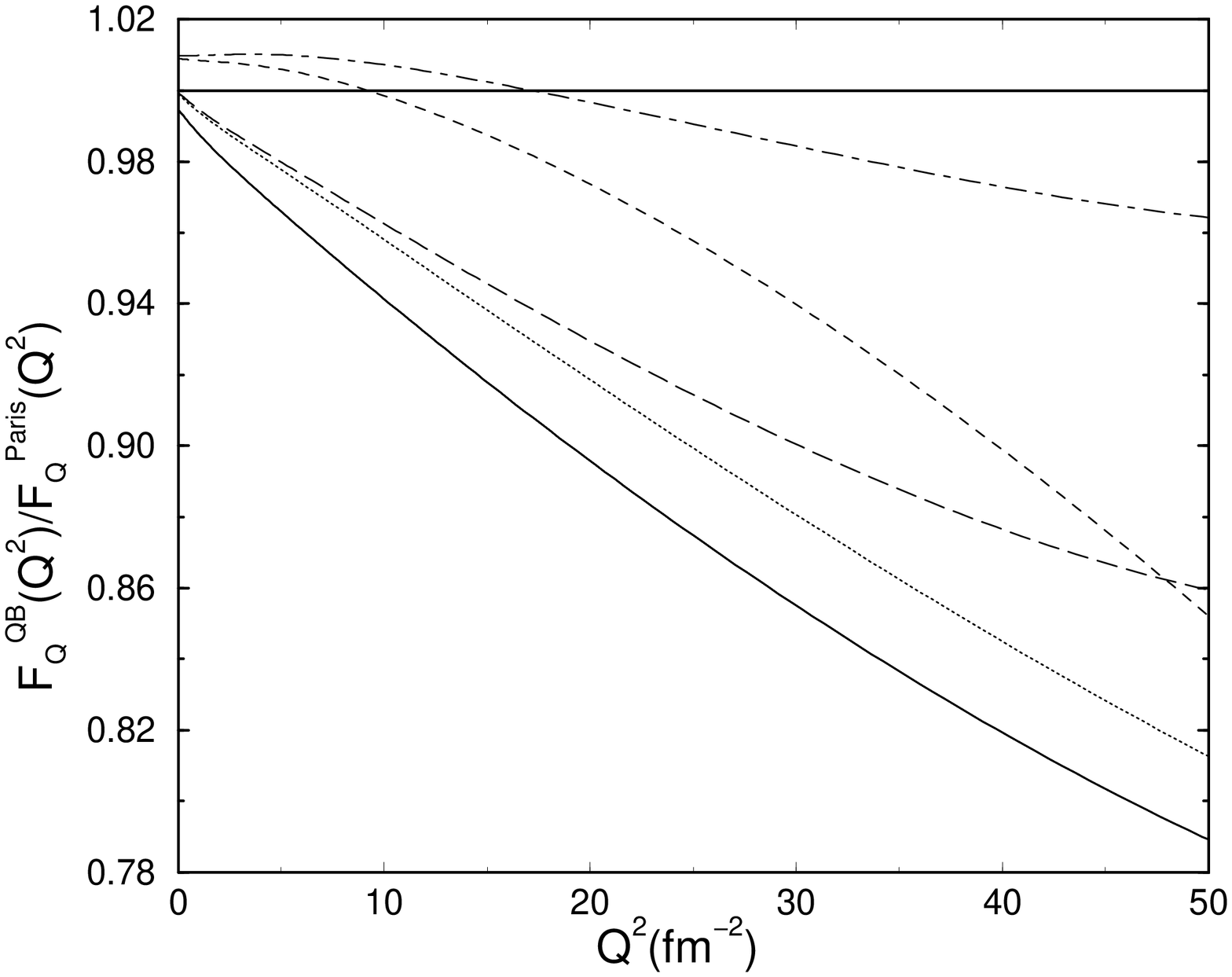, angle=270, width=6cm} \hspace{1cm}
\epsfig{ file=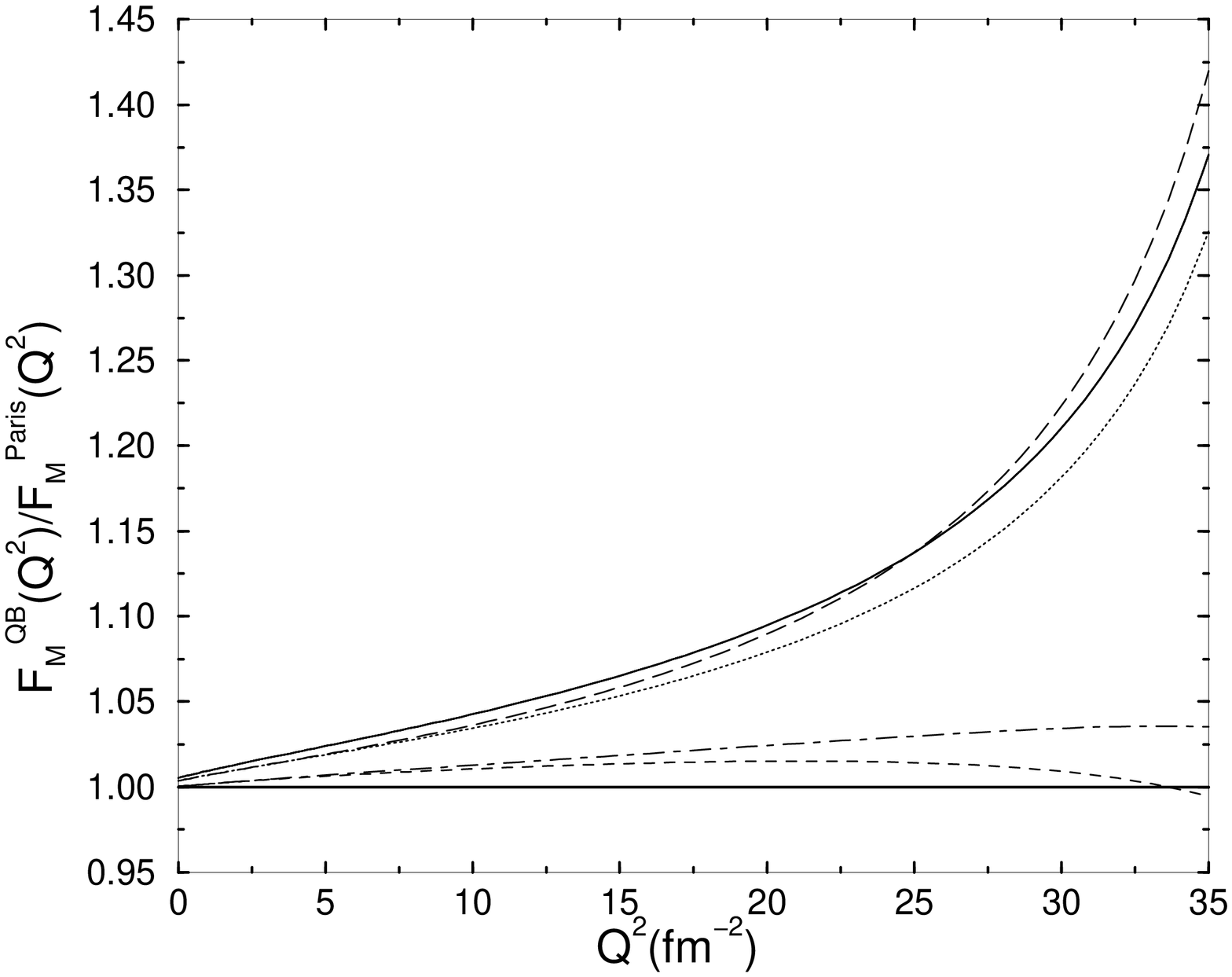, angle=270, width=6cm}}  
\end{center}
\caption{Effect of a non-local term, $\{ {\bf p^2}/M,\,W_T\,S_{12}\}$ 
together with 
off-shell effects for the quadrupole and magnetic form factors:  Curves 
represent the ratio of the Bonn-QB to the Paris quadrupole and magnetic form 
factors (left and right parts respectively). They are defined as in Fig. 
\ref{figx2}.}
\label{figy2}
\end{figure} 

\begin{figure}[htb!]
\begin{center}
\mbox{ \epsfig{ file=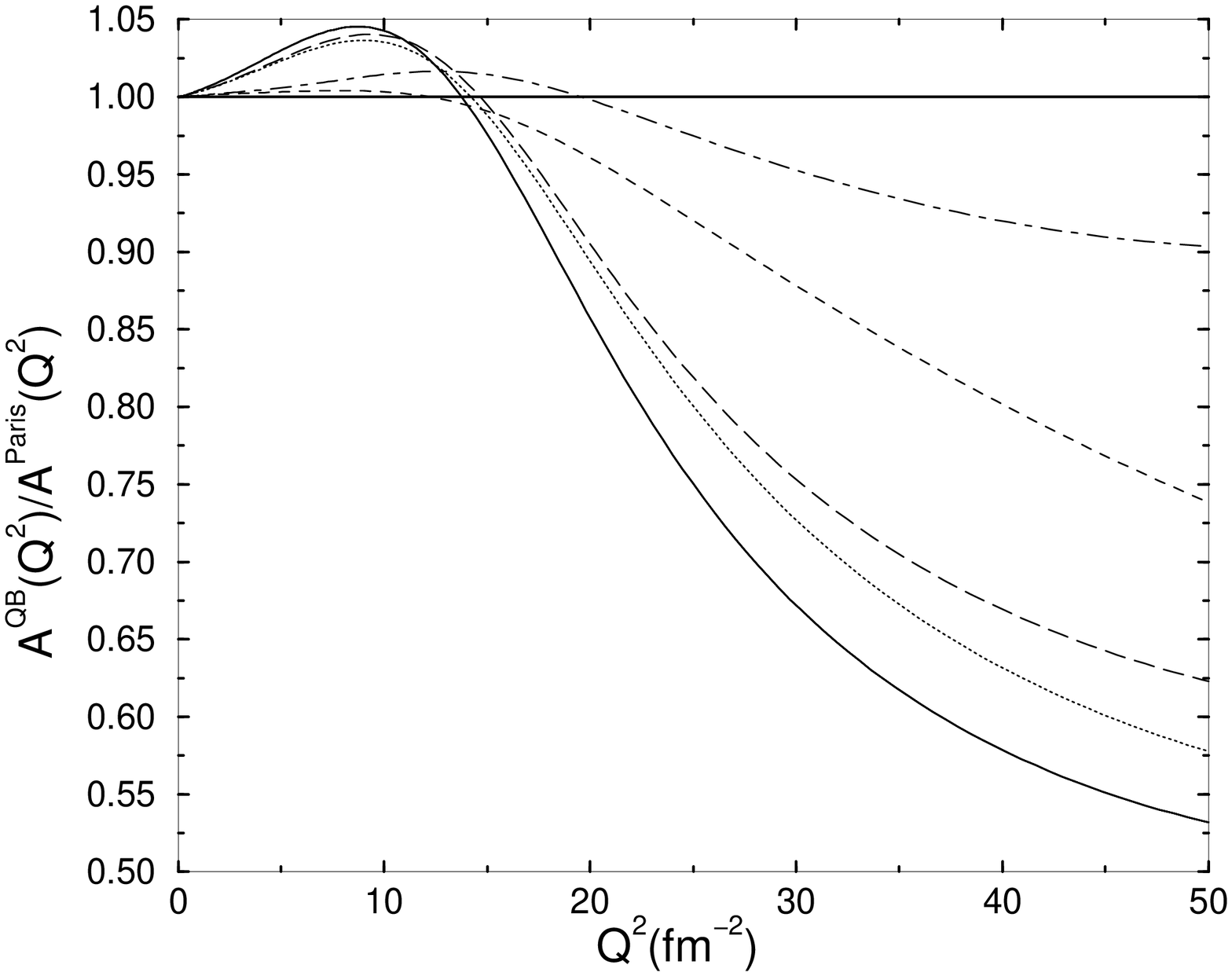, angle=270, width=6cm} \hspace{1cm} 
\epsfig{ file=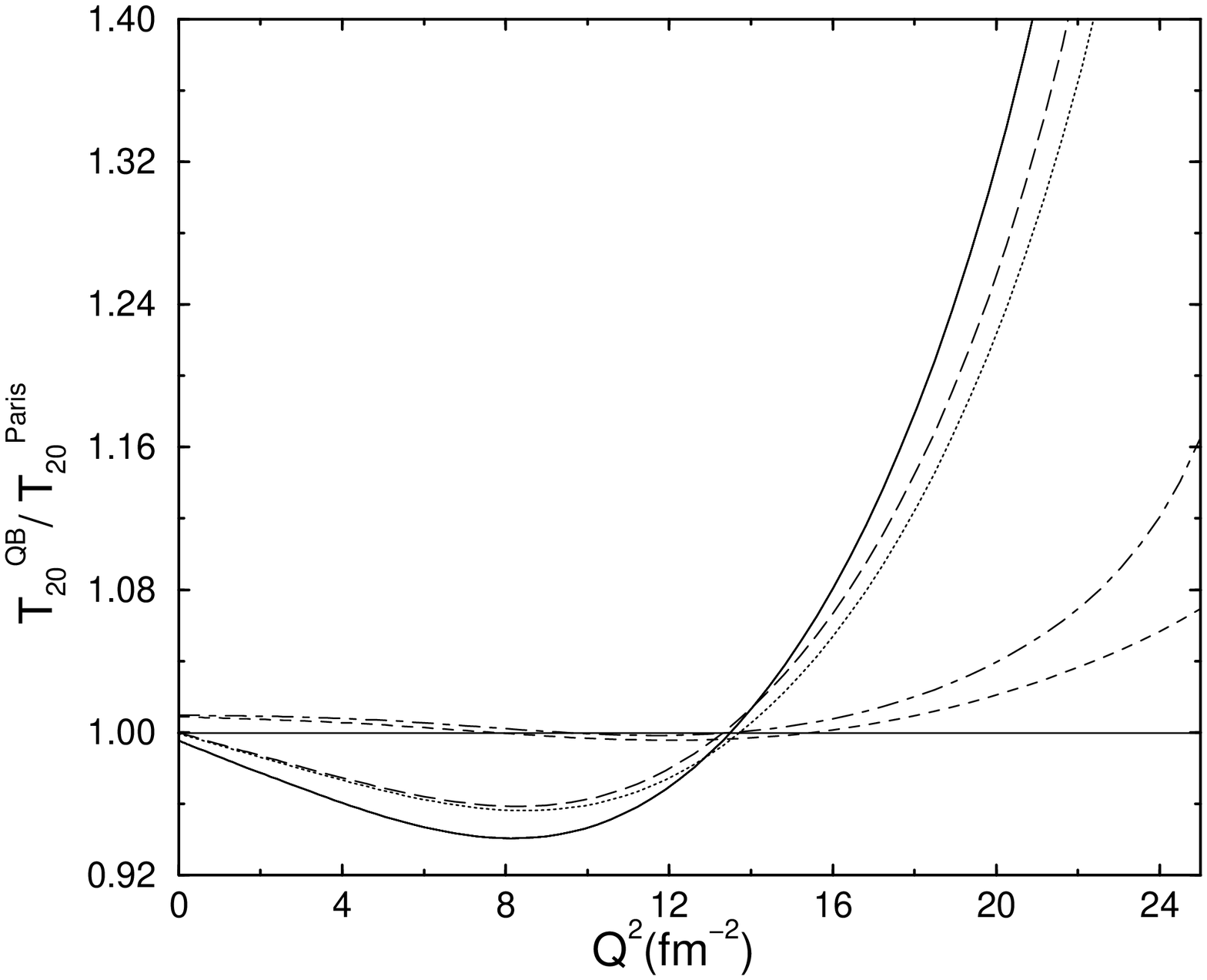, angle=270, width=6cm}}  
\end{center}
\caption{Effect of a non-local term, $\{ {\bf p^2}/M,\,W_T \,S_{12}\}$ 
together with 
off-shell effects  for $A(Q^2)$ and $T_{20}(Q^2)$ observables:  Curves represent 
the ratio of the Bonn-QB to the Paris $A(Q^2)$ and $T_{20}(Q^2)$ observables. 
They are defined as in Fig. \ref{figx2}.}
\label{figAT202}
\end{figure} 

\begin{figure}[htb]
\begin{center}
\mbox{ \epsfig{ file=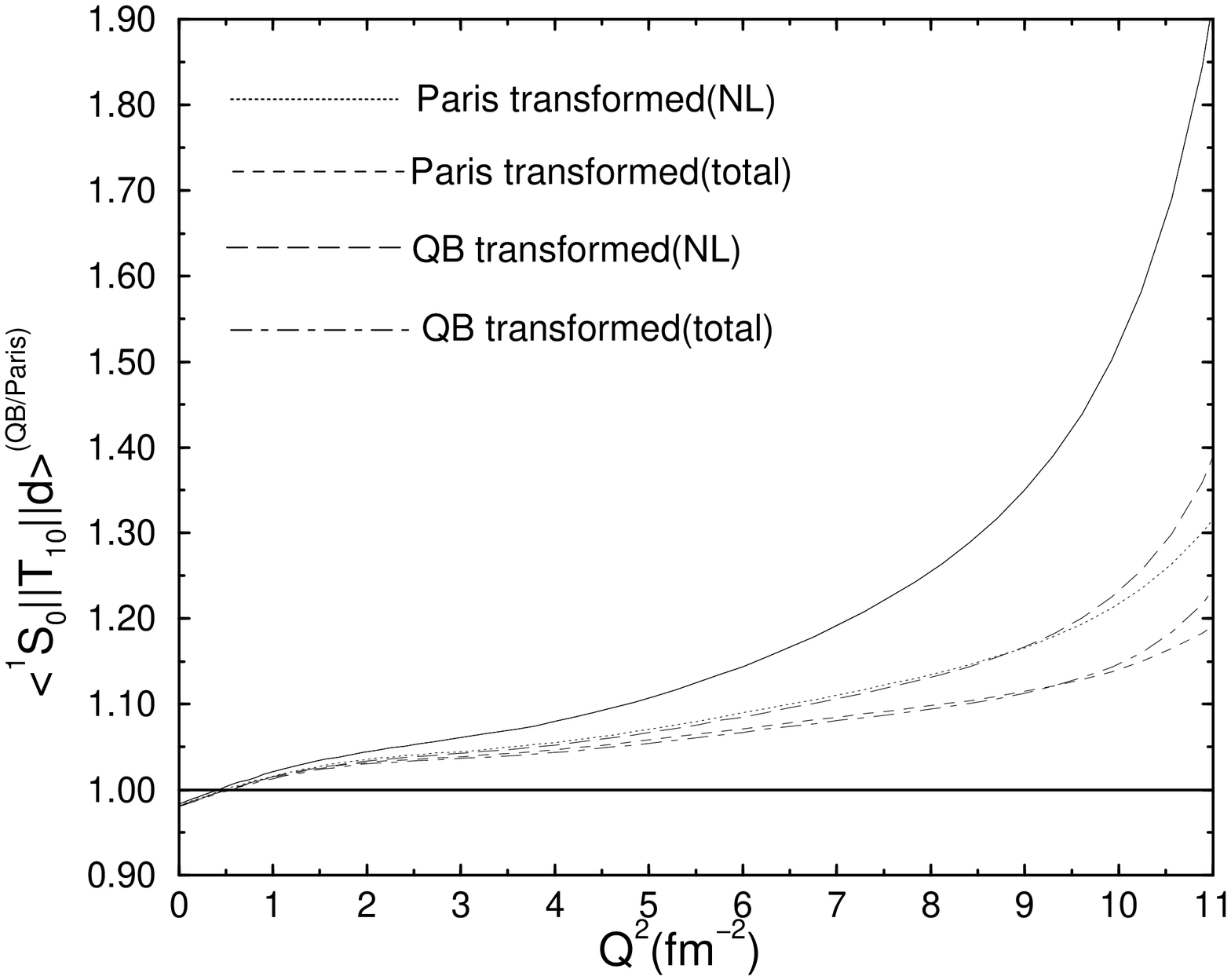, angle=270, width=6cm} \hspace{1cm}
\epsfig{ file=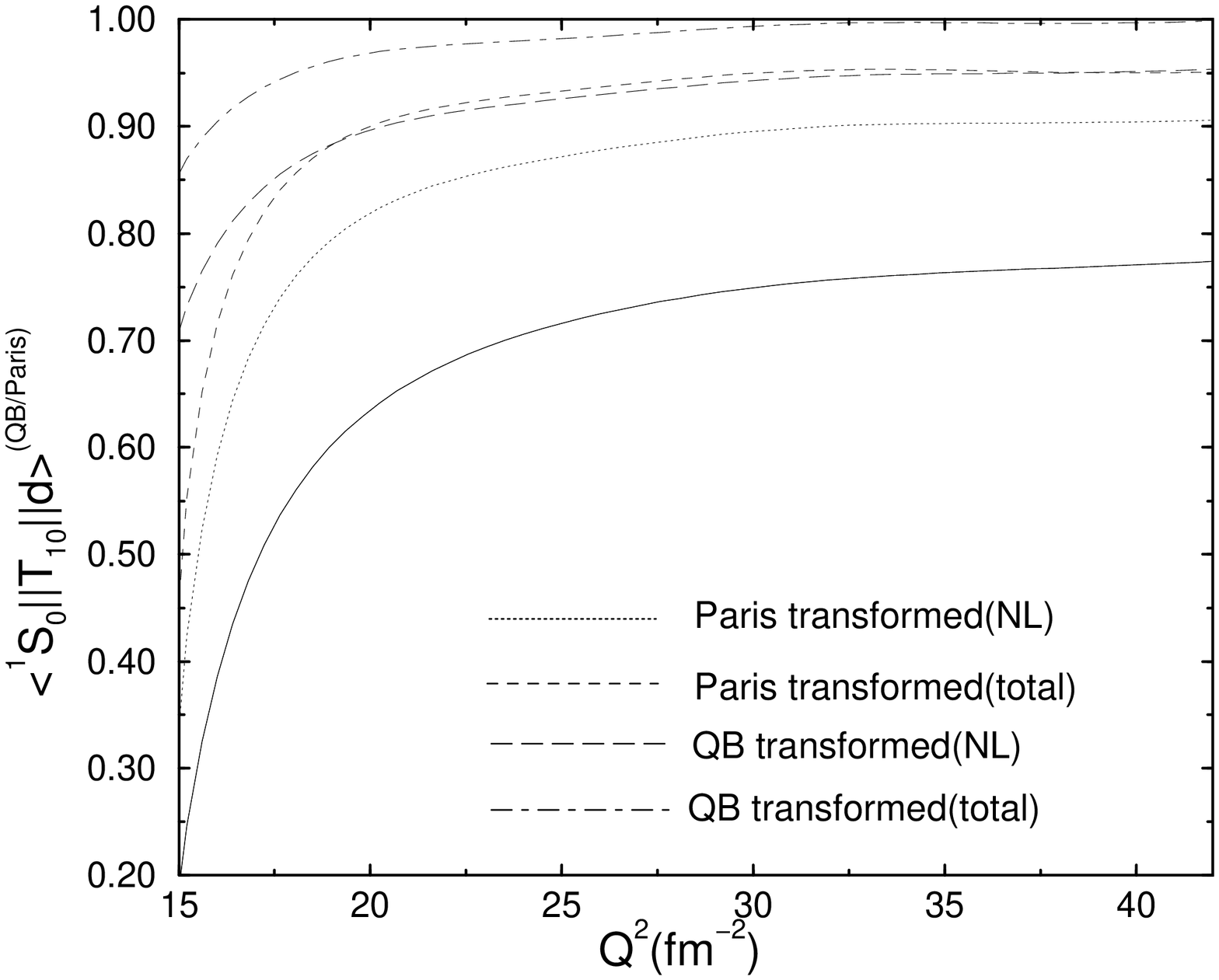, angle=270, width=6cm}}  
\end{center}
\caption{Effect of a non-local term, $\{ {\bf p^2}/M,\,W_T \,S_{12}\}$ 
together with 
off-shell effects for the deuteron electrodisintegration form factor near 
threshold: Curves represent the ratio of 
the Bonn-QB to the Paris form factor from  $Q^2=0$ to $Q^2=11\,{\rm 
fm}^{-2}$ for the left part and  from  $Q^2=15$ to $Q^2=40\,{\rm fm}^{-2}$ 
for the right part. The ratio of the uncorrected predictions is given by the 
continuous line. The ratio when the Paris (Bonn-QB) w.f. is transformed for 
the first ( $\{ {\bf p^2}/M,\,W_T \,S_{12} \}$) and second (off-shell) 
non-locality is 
given by the dotted and short-dashed (long-dashed and dash-dotted) lines.}
\label{fig1S0}
\end{figure} 

The effect of a tensor term like $\{ {\bf p^2}/M,\,W_T\,S_{12}\}$ on 
electromagnetic 
observables is shown in three figures for the case of the Bonn-QB and 
Paris models, which differ by such a term. These models also differ by 
terms that have an off-shell character and were studied in an earlier work. 
In presenting the various effects, we first present the new one considered 
here and then add the older one.

The effect of the non-local tensor term on the charge form factor is 
presented in Fig. \ref{figx2}. It explains a little part of the difference 
between the Bonn-QB and Paris model predictions. By far, the dominant 
effect is due to the off-shell effect associated to the $\pi$- and 
$\rho$-exchange tensor force. Quite similar results hold for the quadrupole 
and magnetic form factors shown in Fig. \ref{figy2} and, as a consequence, 
for the structure function, $A(Q^2)$, and the tensor polarization, 
$T_{20}(Q^2)$, 
both given in Fig. \ref{figAT202}. In these two cases, possible discrepancy is 
completely removed below $15\,{\rm fm}^{-2}$. There may be still some beyond in 
relation with that one observed for the quadrupole form factor in Fig. 
\ref{figx2}. We notice that in dealing with the non-local tensor force, we 
omitted in the Bonn-QB model a term of the form:
\begin{equation}
\sigma_1 \ccdot {\bf p'}\times {\bf p}\;\;
\sigma_2 \ccdot {\bf p'} \times {\bf p} \; .
\label{SLSL}
\end{equation}
This one appears at the fourth order in $p/M$ and, moreover, arising from 
vector-meson exchange,  has a short-range 
behavior. In configuration space, this term gives rise to a $\vec{\sigma_1} 
\cdot \vec{L}  \;\; \vec{\sigma_2} \cdot \vec{L} $ term, which is present 
in the Paris model considered in this work, non-local terms that are similar 
to those we discussed  and another non-local one, proportional to 
$\vec{\sigma_1} \ccdot \vec{p}  \;\; \vec{\sigma_2} \ccdot \vec{p} $
\cite{STOK}. This last one could be dealt with by using a 
transformation identical to that given in Eq. (\ref{2e11}), but with different 
relations of $V_1$ and $V_2$ to the potential strength (see appendix). At first 
sight, this new term could contribute in the right direction. However, in 
discussing its contribution, we should also include higher order contributions 
due to eliminating non-local terms at the lowest order, $[S,\;V_{NL}]$. These 
induced terms have an extra factor $1/M^2$ (for the tensor part) but a range 
quite similar to that of vector-meson exchange. 

Most results presented above concern an elastic transition. Some 
have been obtained for an inelastic transition, namely the deuteron 
electro-disintegration near threshold which involves a matrix element between 
the deuteron $^3S_1\;(+^3D_1)$ state and the $^1S_0$ scattering state. They are 
presented in Fig. \ref{fig1S0}. Examination of these results shows features 
quite similar to the ones observed in other cases. In principle, this process 
allows one to get insight on the effect of a non-local term in a channel, 
$^1S_0$, different from the deuteron one. However, as most of the effect is 
related to the tensor force, the conclusion one can draw from this case is 
limited.

As for the case of the non-local scalar term considered in the previous 
subsection, we looked at other models (Bonn-QB replaced by Bonn-CD ; Paris 
replaced by Nij1 or Nij93). Again, a better agreement is obtained with models 
that better fit NN scattering data. In the detail, a few points deserve to be 
discussed however.

\subsection{A few peculiarities}
The studies presented in this work were originally performed with the idea that 
various NN interaction models could be equivalent (up to a unitary 
transformation). With this respect, it is interesting that we can explain a 
large part of the differences between models. The question now arises to 
determine the limitations, if any. By construction, the models are not 
identical. Besides a difference in their non-locality, discussed throughout this 
work, there may be differences in the choice of various ingredients: coupling 
constants, vertex form factors, exchanged mesons together with their mass,  
quality of the fit to NN scattering data, number of free parameters, 
etc....After the non-locality effects have been removed, the comparison of model 
predictions can thus reveal more clearly features in relation with these 
ingredients. We here consider some of them. They involve a few percent effects 
that may not be significant at first sight but become so when appearing 
repeatedly. They also mainly concern the low momentum transfers, the high 
momentum transfer range being too sensitive to a variety of ingredients to make 
firm statements.

\begin{figure}[htb!]
\begin{center}
\mbox{ \epsfig{ file=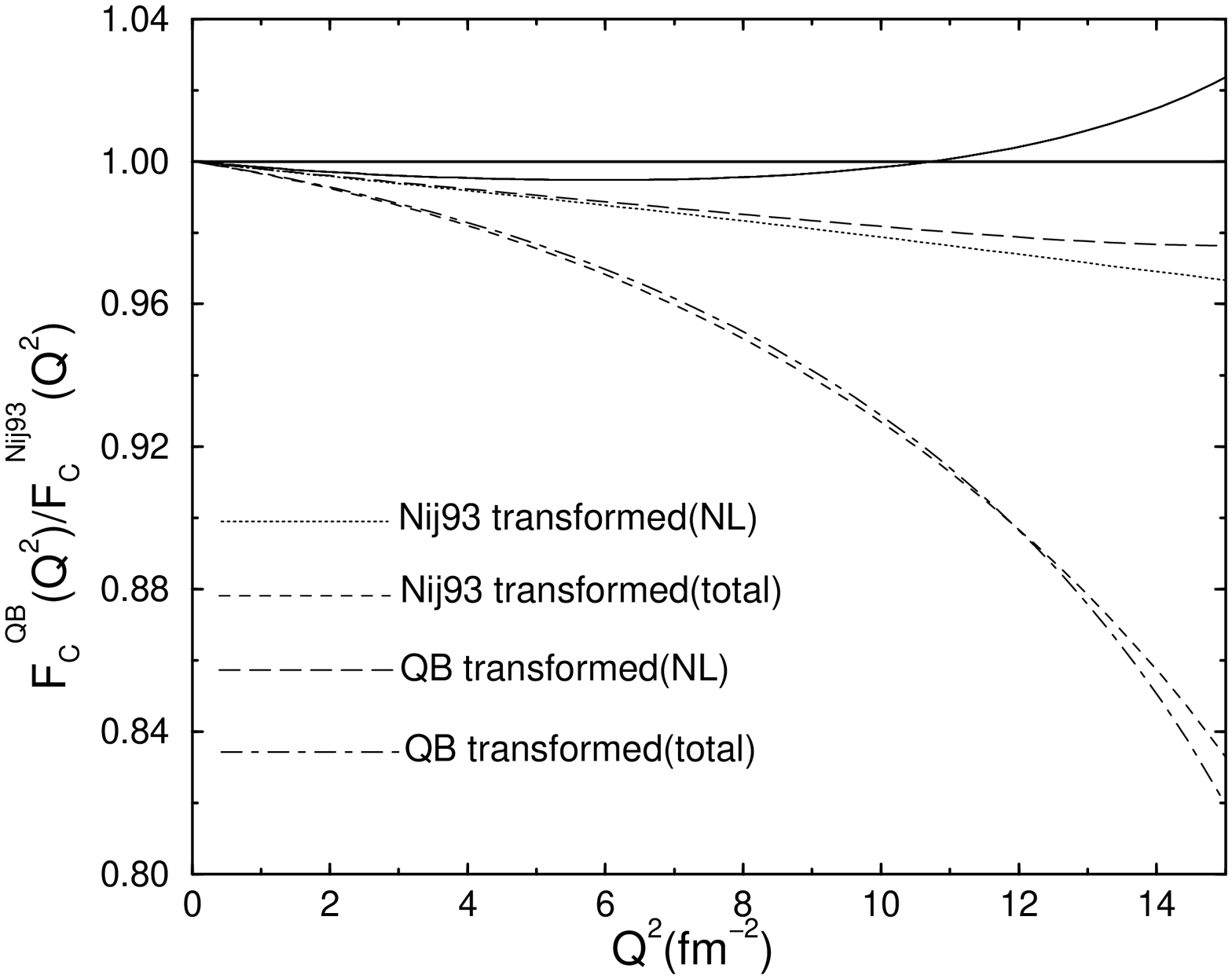, angle=270, width=6cm} \hspace{1cm} 
\epsfig{ file=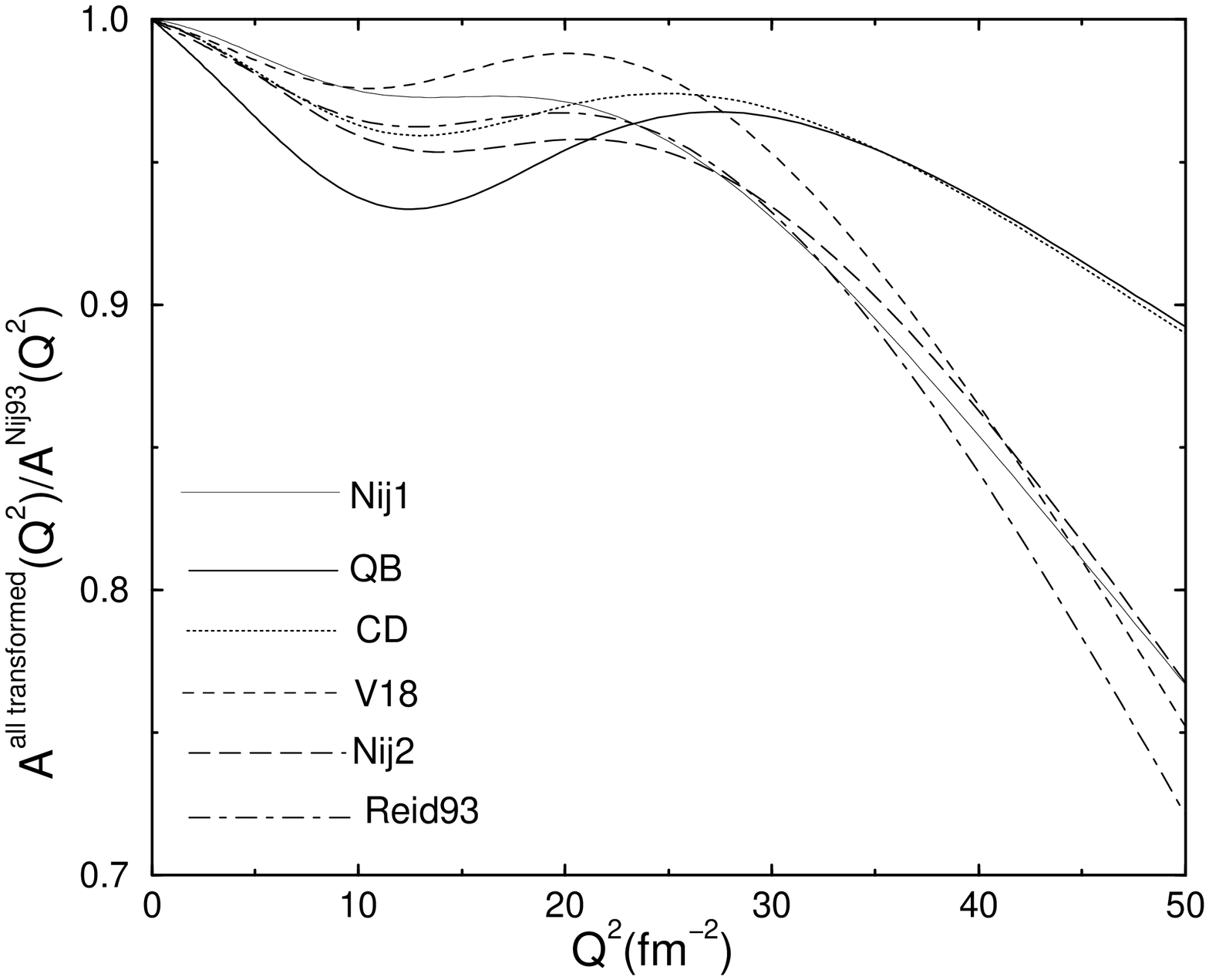, angle=270, width=6cm}}  
\end{center}
\caption{Comparison to results involving the Nijmegen 93 interaction model, for 
the charge form factor (left part) and the $A(Q^2)$ structure function (right 
part).}
\label{fignij3}
\end{figure} 

In Fig. \ref{fignij3} (left part), we show an example where an approximate 
agreement between two models turns into a discrepancy when the influence 
of the non-local 
interaction is accounted for. It concerns the charge form factor calculated with 
the Bonn-QB and Nij93 models in the range  $Q^2=0$ to $Q^2=15\,{\rm fm}^{-2}$. 
Being alone, this example, which is given to illustrate our purpose, does not 
allow one to draw conclusions on one of the two models under consideration. 
However, after incorporating the non-locality effect, it appears that the charge 
form factor for the Nij93 model evidences the same pattern as for the other 
models. This can be seen in the comparison of the $A(Q^2)$ structure function 
calculated with the Nij93 model and with the set of models Nij1, Nij2, Reid93, 
Argonne V18 which, as already mentioned, provide very close results when 
non-locality effects are taken into account. The ratio of these quantities, 
shown in Fig.  \ref{fignij3} (right part), evidences a difference in the slope 
or even a bump in 
the range $Q^2=0$ to $Q^2=15\,{\rm fm}^{-2}$. This tends to suggest that the 
Nij93 model involves specific features that are absent in the other models. It 
is 
tempting to make a correlation with the fact that this model predicts a smaller 
radius than the other ones, whose effect could not be seen clearly without 
removing effects of the non-locality. As to the origin of this smaller radius, 
it could be due to the poorer quality of the fit to scattering NN data (only 15 
parameters). Not independently, it could also be due to the presence in this 
model of a pomeron exchange contribution which is ignored in all the other 
models. Its long range is likely to produce specific features that could not be 
easily accounted for with shorter range contributions to the potential.

\begin{figure}[htb!]
\begin{center}
\mbox{ \epsfig{ file=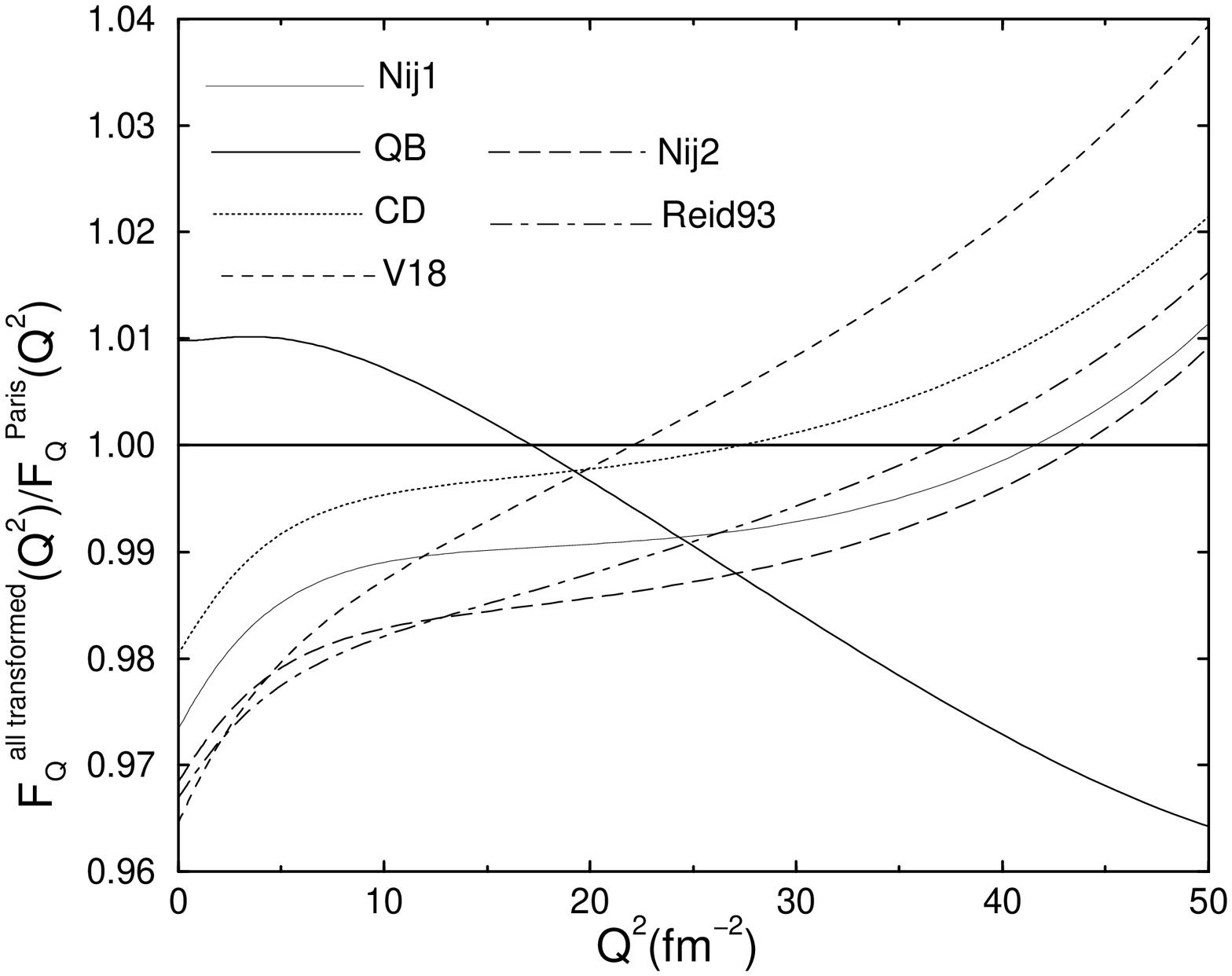, angle=270, width=6cm} \hspace{1cm} 
\epsfig{ file=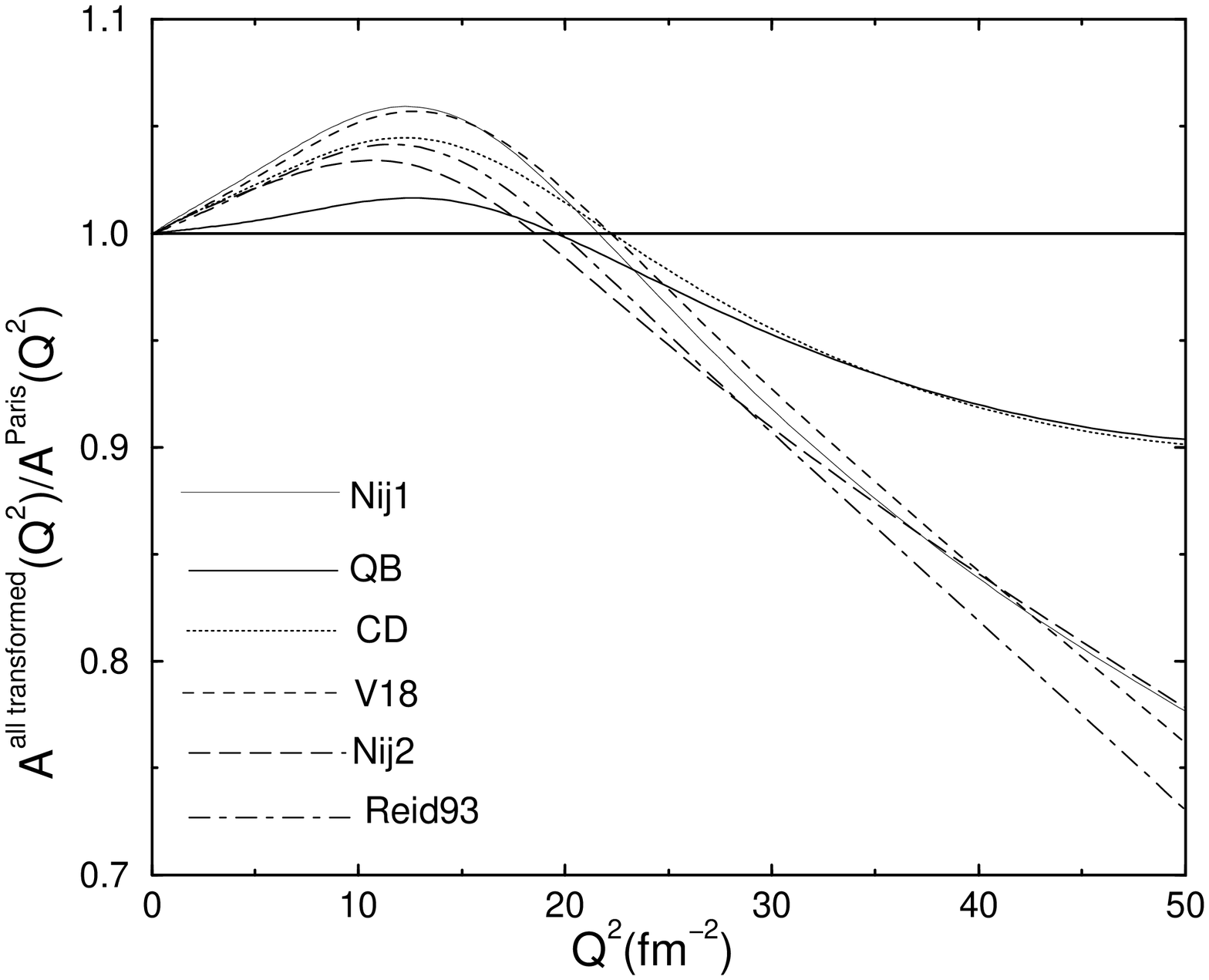, angle=270, width=6cm}}  
\end{center}
\caption{Comparison to results involving the Paris interaction model, for the 
quadrupole form factor (left part) and the $A(Q^2)$ structure function (right 
part).}
\label{figpari}
\end{figure} 

Another particular feature shows up when comparing the $A(Q^2)$ structure 
function calculated with the Paris model to that one calculated with 
one of models Nij1, Nij2, Reid93, Argonne V18 and Bonn-CD and the appropriate 
corrections for non-locality effects. The ratio, which is shown in 
Fig. \ref{figpari} (right part), evidences a slight bump  in the 
range $Q^2=0$ to $Q^2=15\,{\rm fm}^{-2}$. A similar but less pronounced 
effect is seen with the Bonn-QB model. Again, the difference in the slope 
suggests to look at the deuteron radius predicted by the models. It is 
thus found that the radius for the Paris and Bonn-QB models, after 
accounting for  non-locality effects,  are slightly larger than for 
the other models. Before wondering about a possible origin for this 
difference in the radius, we emphasize another feature that may have 
some relation. It concerns the quadrupole form factor and more specifically 
the quadrupole moment. In many cases, we could explain differences in the 
quadrupole moment by the difference in the non-locality of the models. 
This is not the case for the Paris model (and to a lesser extent for 
the Bonn-QB model) as it can be seen in Fig. \ref{figpari} (left 
part). The quadrupole moment for the Paris model is always larger than 
for the other models by an amount of 3-4\%. There is no direct relation 
to the radius discussed above although a larger value will certainly 
contribute to produce a larger quadrupole moment. An explanation could 
be looked for in the value of the $\pi NN $ coupling constant which 
is larger for the Paris (and Bonn-QB) model than for the other models 
which, for some of them, have been built on the prejudice of a value 
around $g^2_{\pi NN} /4\pi\simeq 13.5$ \cite{DESWART}. The value for 
the Paris model is closer to that one advocated by Ericson et al., 
$g^2_{\pi NN} /4\pi\simeq 14.11$ \cite{ERIC}. 
A larger coupling implies a larger attraction at large distances 
and therefore a larger probability to  find the neutron and the proton 
far apart in the deuteron. In the case of the Paris model, one can add 
that the explicit two-pion exchange contribution is likely to also produce 
a larger attraction at long distances in comparison with an approximation 
where this contribution is treated as an exchange of an effective $\sigma$ 
with mass around 500 MeV.

\section{Conclusion}
In this work, we have studied the effect of non-local terms on the 
description of the NN interaction. We especially looked at terms which 
involve the anticommutator with the operator $p^2$, namely 
$\{ {\bf p^2}/M,\tilde{ W}_{S,T}\}$. The present work completes another one 
where 
terms of the commutator form $[ {\bf p^2}/M, i\,U]$, with an off-shell  
character, were considered. In dealing with these terms, we relied on 
an explicit form for the transformation allowing to get rid of them, 
which furthermore was treated at the first order in the interaction. 
In comparing models, this approach is probably not as good as an approach 
based on inverse scattering methods. It nevertheless has the advantage 
to be based on the explicit identification of non-local terms in 
interaction models and to provide an insight on the currents that have 
to be associated with the models. As far as we can see, the method 
is accurate enough for our purpose, especially in view of its qualitative 
character for a part. On the other hand, the study has been extended 
to other interaction models, allowing one to draw conclusions that the 
comparison of two models only could not. 

For our study of the newly considered terms, we took the scalar part, 
$\tilde{W}_S$, from the Paris model while the tensor part, $\tilde{W}_T$, 
was taken from the sum 
of a contribution due to pion exchange, rather well determined, and 
a short range contribution. It is reminded that there is no a priori 
guarantee that these non-local terms can explain differences in 
interaction models which could have other sources and in some cases 
could not be explained (when they involve differences in observables 
such the asymptotic normalizations, $A_S$ and $A_D$). 

We found that the scalar part, $\tilde{W}_S$,  has quite small 
contributions 
to static quantities, such as the charge radius or the quadrupole moment. 
To evidence them, one has to get rid of contributions due to different 
asymptotic normalizations for instance, which compete with the effects 
we are looking at or are even larger. Some sizeable effect however shows 
up for dynamical observables such as form factors. It is difficult 
to make a definitive statement on the size of the effect in view 
of various uncertainties, but its sign and its magnitude can reasonably 
explain differences between models differing by the above term. 

Concerning the tensor part,  $\tilde{W}_T $, the 
contribution to static properties is more important. It, in particular, 
explains a non-negligeable 
part of the difference in the deuteron D-state probabilities between 
the Bonn models, which account for the full structure of Dirac spinors, 
and the other ones. When added to the effect of non-local terms with an 
off-shell character, which is larger, not much difference remains to be 
explained. For dynamical observables, the effect of the above term 
goes also in the right direction and accounts for a half of what could 
not be previously explained with the off-shell term. It would be 
interesting to see what  would do an extra non-local tensor term 
omitted in our analysis. This one has a short range behavior but also 
a different structure which would require further theoretical elaboration. 

While a large part of differences evidenced by the comparison of models 
with various non-localities can be ascribed to this feature, we would 
not like to give the impression that every difference can be explained 
that way. Interestingly, after removing the above effects, some differences 
between models, that were masked until now, more clearly appear. 
The correlation between the slope of the deuteron structure function, 
$A(Q^2)$, and the deuteron radius, which is expected to hold in any case 
at very small $Q^2$, is found to extend to higher $Q^2$, up to 
$10\,{\rm fm}^{-2}$. Knowing that there is for local models an empirical 
relation between the radius and the asymptotic normalization, $A_S$, 
the above observations raises the question of which model is the best 
with this respect, taking into account that this quantity is not changed 
in the transformations we performed. 

The examination of the  quadrupole form factor, especially at $Q^2=0$, 
where it identifies to the quadrupole moment, $Q_D$, also shows interesting 
features. For this one, contributions involving the non-locality can 
amount to 1\%, which is less than the difference between the upper value 
obtained with the Paris model (0.279 fm$^2$) and the lower ones (around 
0.270-0.271 fm$^2$), obtained with the Nij1, Nij2, Reid93, Bonn-CD models. 
Curiously, there is no direct relation to the value of the ratio, 
$\eta=A_D/A_S$, which mainly determines this moment. After the effect 
of the non-locality is accounted for, which mainly affects values  
obtained with the Bonn models with respect to the other ones, we observe 
that the quadrupole moments are exactly in the same order as the $\eta$ 
values and not far to be proportional to them. Accounting for the 
non-locality effects therefore restores some regularity that was 
lost when considering predictions of Bonn models.

Throughout this paper, we ignored comparison with measurements, being 
essentially concerned by the comparison of theoretical predictions. 
We believe it is important in discriminating between models to separate 
effects that are due simply to a mathematical representation of the 
interaction, with more or less non-locality, and effects that have 
their origin in genuine aspects of the models. In making a comparison 
to measurements, one has first to determine among the various approximately 
phase-shift equivalent descriptions of the NN interaction which one 
is the more realistic, in the sense of relying on degrees of freedom 
that are as little effective as possible. Then, one has to also 
account for contributions from two-body currents corresponding 
to this optimal model. Some work along these lines has recently 
been done with the aim to extract the neutron charge form factor from the 
knowledge of the deuteron quadrupole form factor \cite{SICK}.

We are very grateful to R. Machleidt, C. Elster and R. Schiavella 
for providing us with material that has been employed here 
or has been useful in analyzing our own results. \\

{\bf APPENDIX} \vspace{5mm} \\
{\bf Induced contributions to the interaction, $\Delta V$} \\ 
When the non-local terms of the form $\{ {\bf p^2}/M, \tilde{W}\}$ are 
removed from an interaction model to transform it into a local potential, 
local terms of the same order in the interaction are generated. 
Their expression is given here for the scalar and tensor parts. 
For the first one, there is  a factor 2 change with what we give 
in Ref. \cite{AMGH1}. The new convention now agrees with the  
definition of the non-local term in Ref. \cite{PARI}. For the 
second one, we give an expression which evidences its tensor character. 
Due to relations between $V_0$ and $W_S$, or $V_1$, $V_2$ and $W_T$ 
for the scalar and tensor non-local terms respectively 
(see Eqs. \ref{2e9}, and \ref{2e12}, \ref{2e13}), other expressions 
could be obtained. Those given below are among the simplest we could find:
\begin{eqnarray}
\Delta V_S^0=-\frac{1}{2M}\;W_S'' 
+\frac{1}{M} \; \Big\{ L^2 \; , \;\frac{V_0'}{r}  \Big\},
\label{appS}
\end{eqnarray}

\begin{eqnarray}
\Delta V_T^0 &=& \frac{2}{M}\,\Big\{ S_{12}(\vec{L}) \, ,\;\frac{V_2}{r^2}\Big\}
+\frac{1}{M}\;  \Big\{ L^2 , \,S_{12}(\hat{r})\;r\;
\Big(\frac{V_1}{r^2}\Big)'   \Big\} \nonumber \\
& & -\frac{1}{2\,M}\;S_{12}(\hat{r})\;\Big( W_T''+12\,\frac{V_2}{r^2}\Big),
\end{eqnarray}
where
\begin{equation}
S_{12}(\vec{A})= \frac{1}{2}\;
(\vec{\sigma}_1\ccdot \vec{A}\;\;\vec{\sigma}_2\ccdot \vec{A}
+\vec{\sigma}_2\ccdot \vec{A}\;\;\vec{\sigma}_1\ccdot \vec{A})
-\frac{1}{3}\vec{\sigma}_1\ccdot\vec{\sigma}_2\; \vec{A}^2,
\label{appT}
\end{equation}
with $\vec{A}=\vec{L}$ or  $\vec{A}=\hat{r}$. 

As can be noticed, removing non-local terms introduces terms dependent 
on the total angular momentum, in agreement with the expectation that 
the phase-shift equivalent potential is local, wave by wave however. This 
is especially transparent for the scalar part. It is also noticed 
that the global range is not affected, contrary to the removing of 
off-shell effects, but the potential is somewhat more singular 
due to the presence of derivatives of $V_0,\;V_1,\;V_2$, or $W_S$ and $W_T$ 
in the above expressions. Particular caution may be therefore required 
in making qualitative statements concerning the effect of these terms. 
For instance, in case where $W_S$ would be given by a function, 
${\rm exp} (-\mu\;r)$, the term involving the second derivative in Eq. 
(\ref{appS}) would produce a contribution that is attractive 
at large distances. However, it also produces a repulsive $\delta(r)$ 
term that will show up at very short distances. This one,  
often discarded, can give rise to the suppression of the S-state 
wave function that local models often evidence in this range. Although the 
singular character of the correction to the  interaction does not necessarily 
prevent one to perform meaningful calculations, we felt better to cut-off the 
short-range part of the non-local interaction. In this way, we perhaps 
underestimate some of the effects but those we calculate are thus determined 
by the longer range part of the force, which is physically better known.\\ \\ 

{\bf Contributions to the static moments}\\
We here give the corrections to the static moments that can be obtained a direct 
calculation ($P_D$) or from Eqs. (\ref{4e27}) for the form factors in the limit 
$Q \rightarrow 0$ ($\Delta <r^2>$, $\Delta Q$).
\begin{eqnarray}
\Delta
P_D=\frac{\sqrt{8}}{3}\int_0^\infty dr \;
\Bigg(\Big(W_T(r)-6\,V_2(r)\Big)\;u(r)\;w(r) \hspace{1.6cm}
\nonumber \\
+2\,r\;\Big(V_1(r)+2\,V_2(r)\Big)\;u'(r)
\; w(r))\Bigg),
\end{eqnarray}
\begin{equation}
\Delta <r^2>_S=\frac{1}{2}\int_0^\infty dr\;
r^2\;V_0(r)\;\Big(u^2(r)+w^2(r)\Big),\hspace{3.3cm}
\end{equation}
\begin{eqnarray}
\Delta <r^2>_T=\frac{\sqrt{8}}{3}\int_0^\infty dr\;r^2\;
\Big(V_1(r)+2\,V_2(r)\Big)\; \hspace{3cm}
\nonumber \\ 
\times \Big(u(r)\;w(r)-\frac{1}{\sqrt{8}}\,w^2(r)\Big),
\hspace{3cm}
\end{eqnarray}
\begin{equation}
\Delta Q_D^S=\frac{\sqrt{2}}{5}\int_0^\infty dr\;
r^2\;V_0(r)\;\Big(u(r)\;w(r)-\frac{1}{\sqrt{8}}\,w^2(r)\Big),\hspace{3.0cm}
\end{equation}
\begin{eqnarray}
\Delta Q_D^T=\frac{2}{5}\int_0^\infty dr\;
r^2\;\Bigg(V_2(r)\Big(w^2(r)-u^2(r) +\frac{1}{\sqrt{2}}\,u(r)\;w(r)\Big) 
\nonumber \hspace{1.6cm}\\ 
-\frac{1}{3}\Big(V_1(r)+2\,V_2(r)\Big)\;
\Big(u^2(r)+\frac{3}{2}\,w^2(r)-\sqrt{2}\,u(r)\;w(r)\Big)\Bigg).
\end{eqnarray}

{\bf About another non-local term}\\
We here consider a non-local term of the form:
\begin{equation}
V_{NL}= 
\Big\{\frac{\vec{\sigma_1} \ccdot \vec{p}  \;\; \vec{\sigma_2} \ccdot 
\vec{p}}{M},\; W\Big\} \;.
\end{equation}
This one may arise from approximating the term given by Eq. (\ref{SLSL}) when 
treated in configuration space (order $1/M^4$). It could also come from the 
exchange of a $a_1$ meson at the order $1/M^2$. It contains both a scalar part, 
which has a form identical to the one discussed in the main text of the paper 
and a tensor part. This last term can be transformed into a local one at the 
lowest order, using the same transformation as for the other non-local tensor 
term (\ref{2e11}). The functions $V_1$  and $V_2$ now fulfill the relation:
\begin{eqnarray}
V'_1(r)+V'_2(r)&=&0,\label{app30} \\ 
V_1(r)+V_2(r)+(r\,V_2(r))'&=&W(r).
\label{app31}
\end{eqnarray}
The above system of equations can be formally solved with the result:
\begin{eqnarray}
 V_2(r)&=&-\frac{1}{r}\int_r^\infty \frac{W(r')}{r'}dr',
\label{app34}  \\
V_1(r)&=&-V_2(r).
\label{app35}
\end{eqnarray}

\end{document}